\documentclass[10pt,a4paper]{article}
\usepackage{graphicx}
\usepackage{tabularx}
\usepackage{multicol}
\usepackage{tabto}
\usepackage{url}
\usepackage{enumitem}
\usepackage{textcomp}
\usepackage{eurosym}
\usepackage{wasysym}
\usepackage{xcolor}
\usepackage[labelfont=bf]{caption}
\usepackage[yyyymmdd]{datetime}

\usepackage{flowchart}
\usetikzlibrary{shapes,arrows.meta,chains,external}
\newcolumntype{L}[1]{>{\raggedright\arraybackslash}p{#1}}
\newcolumntype{C}[1]{>{\centering\arraybackslash}p{#1}}
\newcolumntype{R}[1]{>{\raggedleft\arraybackslash}p{#1}}

\newcommand{\cz}[1]{\textit{\textbf{#1}}}

\begin{document}

\bibliographystyle{plain}
\thispagestyle{empty}

\begin{center}
\large{\bf Response to ``The digital pound:\\
a new form of money for households and businesses?''}\\
\vspace{2em}
\large{\it A consultation response prepared for the Bank of England and HM Treasury}\\
\end{center}
\vspace{0.4em}
\begin{center}
\begin{minipage}[t][][t]{0.32\linewidth}
\begin{center}
\large{\bf Geoffrey Goodell}\\
\vspace{0.2em}
{\texttt{g.goodell@ucl.ac.uk}}
\end{center}
\end{minipage}
\end{center}
\begin{center}
{\large University College London}\\
\vspace{0.4em}
{\large Future of Money Initiative}
\end{center}
\begin{center}
\textit{This Version: \today}
\end{center}

\vspace{4em}

\section{Executive summary}
\label{s:summary}

This document constitutes a response to a Consultation Paper published by the
Bank of England and HM Treasury, ``The digital pound: a new form of money for
households and businesses?''~\cite{boe2023}, the latest document in a
series that includes ``Central Bank Digital Currency: opportunities, challenges
and design''~\cite{boe2020} in 2020 and ``New forms of digital
money''~\cite{boe2021} in 2021.  The Consultation Paper concerns the adoption
of central bank digital currency (CBDC) for retail use in the United Kingdom by
the Bank of England.  We shall address the consultation questions directly in
Section~\ref{s:consultation} of this document.  First, we offer a treatment of
the following specific aspects of the Consultation Paper:

\begin{enumerate}

\item\cz{Motivation and process.}  Overall, although the Consultation Paper
references a variety of specific benefits of CBDC, the Consultation Paper is
broadly silent about the underlying motivation for considering implementing
CBDC in the UK.  Many of the arguments presented throughout the Consultation
Paper depend upon accepting certain assumptions, for example that modern
electronic payments are well-suited to serve the interests of those who use
them, that negate important arguments for researching and developing CBDC,
particularly the need to develop an adequate analogue for cash in the digital
economy.  The lack of clarity about the motivation for developing CBDC
undermines the strength of those arguments and raises doubts about their
validity.

\item\cz{Privacy.} Consumer privacy in the digital economy is non-negotiable.
Many serious articles have identified the threat posed by a cashless society to
the human rights of individual persons, who must inexorably engage in the
economy by making payments as part of their everyday lives.  Reports
acknowledging the threats posed by electronic transactions have been published
around the world, by public-sector
organisations~\cite{allix2019,buttarelli2019,ca-opc2016,edpb2021,cnil2022}, by
private-sector businesses~\cite{mai2019,higgins2021}, and by civil rights
organisations and think tanks across the political
spectrum~\cite{naacp2019,gladstein2021}.  Central banks have also acknowledged
the dangers that cashless payments pose to
privacy~\cite{kahn2018,bindseil2023}, and the degradation of consumer privacy
is also understood to have significant implications for the broader
economy~\cite{garratt2021}.  We reject the argument that those with nothing to
hide have nothing to fear~\cite{coustick-deal2015}, and exceptional access
mechanisms, which seek to provide a means by which authorities can gain
privileged access to data under certain circumstances, have been consistently
dismissed as dangerous and untenable by the security community over the past
quarter century~\cite{abelson1997,abelson2015,benaloh2018}.

Unfortunately, the privacy model for CBDC reflected in the Consultation Paper
wrongly assumes that data protection by trusted parties is a substitute for not
collecting data linking consumers to their transactions in the first
instance~\cite{nissenbaum2017}.  Fortunately, alternative approaches are
available~\cite{rychwalska2021}, and the rise of the digital economy does not
imply that the relinquishment of privacy by individual consumers is inevitable.
Some designs for CBDC that have been proposed promise true and verifiable
privacy for consumers~\cite{bis2022,goodell2022}, and the design of CBDC in the
UK must do so as well.

\item\cz{Custody.} A salient feature of money is the option of individuals to
possess and control it directly, as they can with cash.  However, the custody
model described in the Consultation Paper assumes that consumers will not be able
to possess or control CBDC.  Instead, CBDC would be held in special
``wallets'', which the Consultation Paper describes services offered by providers
who would be \textit{de facto} custodians of the money.  The implicit
definition of the term ``wallet'' as a \textit{service} contrasts with the
internationally accepted definition of a \textit{wallet} as an
``\textit{application or mechanism} to generate, manage, store or use'' digital
assets~\cite{iso22739} (emphasis added).

Direct custody of money as a means to exercise choice is a fundamental right,
and if consumers cannot be in direct possession and control of their money,
then it is not really theirs.  Instead, they would be forced to contend with
the possibility that their use of money might be restricted to certain
purposes, as it was for recipients of the cashless welfare card in
Australia~\cite{dss-au2022}, or curtailed completely, as it was for Canadian
lorry drivers in early 2022~\cite{woolf2022}, who were subsequently exposed to
public blacklisting~\cite{fraser2022}.

Requiring that intermediaries stand between consumers and their own money is
tantamount to requiring that consumers have money within custodial accounts.
We question whether, from the perspective of a consumer, ``digital currency''
residing within an account of this type is significantly different from bank
deposits residing within ordinary bank accounts.  We argue that the promise of
digital currency is intrinsically about the ability for the currency to be
possessed independently of an accounting relationship.

\item\cz{Role of identity.}  Separately from the question of custody, which is
about whether individuals can possess and control their own assets, is the
question of the form that the assets take.  Digital currency assets are
fungible, so in principle they can take the form of either \textit{balances}
(``account-based access'') or \textit{tokens} (``token-based
access'')~\cite{auer2020a}.  Both approaches are technically feasible: Some
cryptocurrencies, such as Bitcoin, conduct transactions using tokens, while
others, such as Ethereum, conduct transactions using balances.

However, the choice of balances versus tokens has implications for the role of
identity in accessing the assets.  In a digital currency system with balances,
a number representing the size of some collection assets must be represented
somewhere, and successive transactions must result in changes to that number.
The state of a balance, therefore, is determined by the net effect of a set of
transactions that are intrinsically associated with that balance and with each
other.  The persistent linkage among transactions implies an \textit{identity}
for the owner of the assets, as the owner of the assets must provide
identification to access the balance to conduct a transaction.  Conversely,
with tokens, all that is needed to conduct a transaction is knowledge of
specific tokens or the cryptographic keys that unlock them.  A token is not
necessarily connected to other tokens, and successive transactions made by the
same person might not be linkable on the ledger.

The independence of a user's transactions is essential for both privacy and
control: Whether or not a particular gatekeeper has custody, the requirement to
use a balance implies that the balance can be locked as a mechanism for
preventing the owner from accessing the assets, or that links among successive
transactions can be forcibly discovered as a means of surveillance, as is the
case with the development of e-CNY system in China~\cite{goodell2021}.
Unfortunately, the design described in the Consultation Paper uses balances
rather than tokens, and the implications for the rights of asset owners are not
explored or justified.

\item\cz{Role of the ledger.} The CBDC design described in the Consultation Paper
makes use of a ``core ledger'' that records the assets held by individuals.
This design has several implications.  Independently of the privacy
considerations and the question of tokens versus balances, the choice to rely
upon the core ledger to record assets in this way implies that all transactions
must be recorded on the core ledger.  This has two problematic consequences.

The first consequence of relying upon the core ledger to store transactions
directly is that users must involve the core ledger in every transaction.
Because a transaction cannot be consummated until the core ledger is consulted,
the scalability of the system is significantly limited.  Transacting parties
must have network connectivity to the core ledger, and they must wait for any
system level delays resulting from other transactions taking place at the same
time.

The second consequence of relying upon the core ledger to store transactions
directly is that the ledger grows as a function of the number of transactions,
which implies that increasing the number of transactions will introduce stress
on the mechanism that maintains the core ledger, creating the opportunity for
denial of service by adversarial actors.  Cryptocurrency systems with this
design, such as Ethereum, typically address this vulnerability by introducing a
transaction charge (also known as a ``gas fee''), which introduces additional
economic frictions as well as fairness considerations related to the question
of whose transactions have precedence and whether there should be implicit
subsidy for users who might otherwise not be able to afford such charges.

A better design would not rely upon the representation of assets (tokens or
balances) directly on the core ledger (distributed or centralised).

\item\cz{Role of the issuer.} The CBDC design described in the Consultation Paper
has an outsized role for the central bank, which not only issues tokens but
also processes transactions, sets the rules, and maintains the historical
record.  This highly centralised design is unlike any general-purpose payment
system that has ever existed before in the United Kingdom.  Cash payments are
naturally decentralised; any two counterparties can consummate a transaction
without involving the issuer by exchanging a physical token.  Modern electronic
payments are also decentralised; they are carried out by networks of
private-sector actors overseen by regulators~\cite{bis2012,bis2012a}.  The
design implied by the Consultation Paper implies that the issuer is required to
process transactions directly, introducing a tremendous operational, technical,
and legal burden upon the central bank.

Trusting the central bank to operate the ledger is fundamentally different from
trusting the central bank to oversee a transaction infrastructure as a
regulator~\cite{goodell2021a}.  Because the role of the central bank described
in the Consultation Paper also includes the singular responsibility for
maintaining the core ledger, there is no technical mechanism to stop this
central actor from changing the historical record.  The best we might expect is
for changes to historical data on the ledger to be observable by periodic
reconciliation activities among third parties with read access to the ledger,
followed by a legal challenge wherein the authoritativeness of the ledger
records would be determined by indexing the intentions of the record maintainer
in some juridical context.  In the absence of a clear procedure for determining
which version of history is true, all parties are forced to accept the state of
the ledger declared by the ledger operator.  Even if we assume that the ledger
operator is not corrupt, all participants in the system must depend upon the
integrity of its agents and the effectiveness of its security practices.
Recent history teaches us that such dependence is a risky
proposition~\cite{tucker2023}.

Finally, when the party that processes transactions also sets the rules, there
is nothing to prevent the rules from changing without warning.  Given how many
retail consumers would rely upon a future retail CBDC infrastructure, this risk
is too much to bear.  In distributed models for the operation of best-execution
networks, such as the National Market System in the United
States~\cite{nms-changes}, the regulator sets the rules but requires
participating private-sector actors to operate the system.  Changes must be
proposed, requested by the regulator, and implemented by all participating
actors before they can take effect.  This procedural approach protects the
system from hazards and is possible only because the system is decentralised in
practice.  Because the system proposed in the Consultation Paper is highly
centralised, it is exposed to risks that decentralised systems can hope to
avoid.

\end{enumerate}

Many of the items in this list are similar to points that were made by
respondents in consultations associated with the earlier Consultation Papers but
were not addressed in the most recent Consultation Paper.  The Bank of England
and His Majesty's Treasury are strongly encouraged to revisit the earlier
consultation responses and consider whether the current approach described in
the most recent Consultation Paper reflects adequate consideration of the points
introduced in those responses.

\section{Detailed response to the Consultation Paper}
\label{s:detailed}

Next, we address some specific items of concern implicit in the Consultation
Paper~\cite{boe2023}, grouped by category:

\newcounter{enumTemp}

\subsection{Motivation and process}

\begin{enumerate}

\item\label{i:cash}\textit{Page 5: ``a digital pound... would sit aside, not
replace, cash''.}  Although the intention is honourable, this objective may be
difficult to achieve in practice, given the decline of cash, which the
Consultation Paper acknowledges as ``being used less frequently by households and
businesses''.  The variable revenues associated with operating cash
infrastructure are rapidly falling below the fixed costs, particularly in the
built environment, wherein cash is commonly refused by merchants already.
Whilst we agree that CBDC should not be seen as a substitute for cash, it can
be viewed as a way to achieve most of the benefits of cash without the
corresponding infrastructure costs.  The orchestration of an appropriate
response to the decline of cash is a key motivation for institutions in the UK,
and indeed for institutions throughout the world, to establish CBDC.

\item\textit{Page 6: ``seeks to begin to build that foundation of public
trust''.} However, the involvement of the general public, including technology
experts and community stakeholders, has been limited to initiatives such as the
CBDC Engagement Forum and CBDC Technology Forum, and there has not yet been a
meaningful public debate about the requirements for CBDC in the UK and policy
implications of the various technical design features.

\item\label{i:schumpeter}\textit{Page 7: ``we will work with the private sector
to explore potential technology solutions''.} Of course, the private sector
should be involved in implementing CBDC and the infrastructure upon which it
relies.  However, involving the private sector in preference to the broader
public during the design phase is a mistake.  The best solution for the British
public, and indeed for the private sector over the long run, might involve a
design that is inconsistent with the (short-term) interests of economic
incumbents, and the process for finding such a solution involves a process that
Schumpeter described as ``creative destruction''~\cite{schumpeter1942}, often
involving forcible disruption to prevailing business models.  Such disruption
is an unlikely consequence of design recommendations promoted by incumbent
private-sector businesses.

\item\label{i:fragmentation}\textit{Page 10: ``If current trends continue...
the monetary system could become fragmented''.}  Fragmentation is already here,
with the decline of cash use.  Some UK vendors refuse cash, and consequently,
they accept only bank deposits, and not central bank money, as a means of
payment.  It is not inconceivable that some vendors have determined that the
costs associated with accepting cash outweigh the benefits, and consumers with
only cash do not have access to goods and services provided by such vendors.
The fact that vendors have the option to refuse cash means that some
transactions do not take place as a consequence of the choice of payment
mechanism alone.  It also means, \textit{de facto}, that vendors can choose to
accept payments only from banks and not directly from retail consumers.

\item\label{i:cash-gap}\textit{Page 10: ``The digital pound would complement
banknotes''.}  See Item \ref{i:cash}.  Note that banknotes no longer serve the
function for which they were designed, leaving a gap in utility for consumers
and businesses.  The digital pound should complement banknotes indeed, but it
should also fill that gap.

\item\label{i:banknote}\textit{Page 13: ``The digital pound would be used like
a digital banknote...  [and would be] useful for everyday payments''.}  This is
an important acknowledgement of the need for digital cash as a means by which
consumers can make retail payments, particularly with the migration of retail
payments to the digital economy.  This need also implies a responsibility to
consider the risk of forfeiture of human rights to privacy, possession, and
control of money that is intrinsic to the methods by which retail consumers
engage with the digital economy, and the need for a public payment
infrastructure that preserves those rights.

\item\label{i:design-phase}\textit{Page 14: ``By the end of the design phase we
will have...  supported business model innovation through knowledge sharing and
collaboration between the private and public sectors''.}  It is imperative that
the process for pursuing engagement with the private sector does not
systematically exclude the voices of the public sector, including other
financial regulators, as well as the general public, including technology
experts, civil society, and community stakeholders, not only in the forthcoming
``design phase'' but also in the process for establishing policy and technology
requirements of the sort that are detailed in the Consultation Paper.  See
Item~\ref{i:schumpeter}.  In particular, we are concerned that the Bank or HM
Treasury might be influenced by designs that support the prevailing business
models of incumbents, or methods that satisfy objectives that have not been
agreed by public consensus, to the detriment of the public interest.  Going
forward, all procurement processes involving design proposals must have full
public transparency and reporting.  But transparency is not enough: To be
equitable, such processes must be matched with corresponding public engagement
on matters of policy and technology design more generally.

\item\label{i:winning}\textit{Page 15: ``Support the development of the broader
UK digital currency technology industry''.} Such support is important, although
the Bank and HM Treasury must be careful to avoid choosing a specific
``winning'' model without full public engagement.  See Items~\ref{i:schumpeter}
and~\ref{i:design-phase}.

\item\label{i:compel}\textit{Page 16: ``After the design phase, there will be a
decision on whether to build a digital pound''.}  The decision about whether to
have a CBDC in the UK should be made in the context of alternatives that can
address the concomitant problems of privacy, possession, and control that
accompany the migration of retail payments to the digital economy.  One
possible alternative is the issuance of stablecoins by private-sector banks;
such stablecoins could be backed by central bank reserves and supported by
mechanisms similar to those that underpin deposit insurance.  Another possible
alternative is legislation to compel retail vendors, including e-commerce
vendors, to accept cash, combined with appropriate procedures to incentivise
the use of cash as needed, including the acceptance of cash by brick-and-mortar
institutions and businesses, perhaps post offices or the offices of wire
transfer service providers, on behalf of Internet-based vendors.

\item\textit{Page 18: ``We will engage stakeholders extensively and be
transparent about our work''.}  The design must not be led by a set of
private-sector ``winners'' that are in a position to set the agenda for
consideration by the CBDC Engagement and Technology Forums and the general
public.  See Item~\ref{i:winning}.

\item\textit{Page 22: ``unbacked cryptoassets do not provide holders with a
safe or stable store of value or a reliable unit of account''.}  Unbacked
cryptoassets are sometimes used as money nonetheless, perhaps because of a
desire for digital cash with cash-like properties of privacy, property, and
custody, and the lack of more suitable alternatives.  Unbacked cryptoassets
could become more popular, and indeed dangerous, if suitable instruments with
cash-like properties are not institutionally supported and regulated.

\item\textit{Page 24: ``Promoting innovation, choice and efficiency in payments
as our payment habits and economy become more digital''.}  The migration of
retail payments to the digital economy also means that an intervention is
required if we desire to affirm and preserve the rights of consumers to
privacy, possession, and control with respect to their payments.  Cash is a
fundamental building block for private property.  In particular, we ask: What
use is money if it inherently belongs to someone else?  See
Item~\ref{i:banknote}.

\item\label{i:cash-role}\textit{Page 24: ``For the digital pound to play the
role that cash plays in anchoring the monetary system''.}  It is certainly
reasonable for CBDC to share its key functional features, including consumer
affordances, with cash.  This choice of language seems to suggest or
acknowledge that there is an expectation that cash will at some point be phased
out.  Despite the insistence by the Bank of England in previous Consultation
Papers that cash will continue to be manufactured, deployed, and used, the
increasing preponderance of vendors in the UK who refuse cash offers a preview
of a possible future scenario wherein cash is seldom used for payments in the
UK.  If the variable revenues associated with accepting cash or operating
components of cash infrastructure fall below the fixed costs, then who will
provide subsidy to ensure that cash continues to be useful?  See
Item~\ref{i:cash}.

For this reason, this sentence highlights a challenge of great importance.
There is a non-negligible chance that we could need CBDC to provide the
functions that cash provides, including not only continuing to ensure that
consumers have verifiable privacy, possession, and control with respect to
their money and how they use it, but also continuing to ensure that consumers
can directly own obligations of the central bank rather than banks that might
fail, a means by which consumers can know in advance that their chosen payment
method will be accepted, and continuing to ensure that money is available for
consumers to use independently of their relationships with private-sector
service providers.  Market fragmentation has already begun.  Establishing a
CBDC might not be the only way to solve this problem, but options are few.  The
time to act is now.

\item\label{i:resilience}\textit{Page 25: ``supporting financial inclusion and
improving domestic payments resilience''.}  We note that both financial
inclusion and payments resilience have been undermined by the migration of
retail payments to the digital economy, since the current digital economy does
not provide a public payment option that preserves the right of consumers to
verifiable privacy, control, and possession of their assets.

\item\textit{Page 25: ``Access to public money... and the uniformity of money
are critical for the smooth functioning of the economy''.}  Certainly true, and
the disappearance of cash undermines both access and uniformity.  See
Item~\ref{i:resilience}.

\item\textit{Page 27: ``The decline in the use of cash is expected to
continue... even though UK authorities are committed to keeping cash available
as long as there is demand for it''.}  It is not obvious that supporting cash
for general-purpose use will be sustainable in the long-term, if the variable
revenues fall below the fixed costs of accepting cash and operating the
infrastructure.  Who will compel merchants to accept cash, and by what
authority?  Who will compel banks to furnish cash to consumers, and by what
authority?  We note the increasing preponderance of ``reverse ATMs'' and
similar mechanisms in foreign cities, such as New York, that require vendors to
accept cash~\cite{tumin2022}; is this a plausible future?  See
Items~\ref{i:cash}, \ref{i:cash-gap}, and~\ref{i:cash-role}.

Although the option to refuse cash is often, \textit{de facto}, exercised in
the UK by vendors who require payments at the point of sale before goods or
services are provided to consumers, certain kinds of payments, for example
restaurant meals, generally take place after the goods or services are
provided.  Can a consumer seeking to settle such an obligation using legal
tender be refused?

\item\textit{Page 27: ``Fragmentation may arise if holders of one form of money
can only interact with others using the same system''.}  Thanks to vendors that
accept only electronic payments, fragmentation is already here.  See
Item~\ref{i:fragmentation}.

\item\label{i:personalisation}\textit{Page 29: ``Recent innovations... have
opened up the payments market... resulting in improved user experience''.}
Recent innovations are at best a mixed blessing; they have also enabled
exclusion (for example, by the refusal of cash) and profiling
(``personalisation'' as a euphemism for surveillance).

\item\label{i:anticompetitive}\textit{Page 31: ``network effects'', ``economies
of scale and scope'', ``data advantages''.}  Control points, anti-competitive
business practices, and surveillance capitalism are hallmarks of the digital
payments industry as it exists today.  Consumers and businesses are not so much
choosing to use digital payment methods as they are being forced to do so
because platforms benefit from data harvesting and the reality of being ``the
only game in town''.

\item\textit{Page 31: ``Th success of a small number of firms can reflect the
fact they offer more innovative products, integration that benefits consumers,
or greater efficiency''.}  This assessment of providers of payment services is
unjustified.  In the case of payments, success is the result of the factors
described in Item~\ref{i:anticompetitive}.

\item\textit{Page 31: ``Big Tech firms' entry in financial services could
benefit many consumers at least in the short-term''.}  This is a matter of
perceived versus real benefits.  Consider, by analogy, the choice of
individuals and businesses to submit to surveillance as a way to reduce
insurance costs.  Although opting for surveillance might indeed reduce costs
relative to resisting surveillance, it is not clear that surveillance
reduces costs to individuals and businesses in the aggregate.  Instead, the
proceeds of market differentiation accrue to the insurance providers.

\item\textit{Page 32: ``the digital pound should not crowd out or prevent other
forms of digital innovation by the private sector''.}  Although enabling
private sector innovation is important, CBDC must not be designed to
intrinsically favour or encourage the use of non-public payment infrastructure
to the exclusion of public payment infrastructure, or to support incumbent
business models.  See Items~\ref{i:schumpeter} and~\ref{i:design-phase}.  It is
also important to acknowledge the dangers of certain private sector payment
options, particularly those that rely upon user surveillance or credit
provision, that involve the exercise of control and introduce negative
externalities.

\item\textit{Page 34: ``UK authorities are committed to ensuring continued
access to cash for those who wish to use it''.}  Access to cash is not enough
to ensure that consumers are able to use it.  Specifically, continued access to
cash is not particularly useful in the absence of assurance that cash will
continue to be accepted.  In the UK, some merchants refuse cash at the point of
sale.  If UK authorities are committed to ensuring access to an effective
public payment option with the properties of cash, then they must ensure that
it will be accepted by vendors.  It might be possible to accomplish this goal
with policy to require brick-and-mortar retail vendors to accept cash and to
compel e-commerce retail vendors to provide a cash payment option.  CBDC might
make such a policy more palatable, if merchants could satisfy the requirement
by accepting CBDC.  See Items~\ref{i:cash} and~\ref{i:compel}.

\item\textit{Page 38: ``the transition in particular could affect some bank
business models''.}  The real issue is that, after years of low interest rates,
some consumer-facing banks have come to depend upon payments, with their
associated transaction fees and data benefits, for a significant share of their
revenue.  These functions are not the same as banking, with questionable
benefit to the public from having banks play this role.

\item\label{i:switch}\textit{Page 38: ``introduction of the digital pound would
result in households and businesses switching some of their bank deposits to
digital pounds''.}  This is true by definition.  However, there is no
justification provided for the implicit assumption that the steady-state
withdrawal and use of digital pounds would be significantly greater than the
steady-state withdrawal and use of cash two decades ago, prior to the migration
of retail payments to the digital economy.  By this token, it might be more
appropriate to view the era of card payments as an anomaly.

\item\label{i:elb}\textit{Page 41: ``effective lower bound (ELB)''.}  As long
as cash exists, the deployment of CBDC cannot lead to a negative ELB.  Even if
cash were to be expunged, a negative ELB would still be unlikely, given the
great prepondernace of alternative assets that can be used as money, such as
foreign cash, cryptocurrencies, precious metals, and so on.

\item\textit{Page 43: ``disintermediation could lead to higher lending rates as
banks experience higher funding costs''.}  It seems unlikely that the economy
has become dependent upon consumers never owning their own money, even for a
brief moment.  See Item~\ref{i:deposits}.

\item\textit{Page 49: ``An important cause of the ELB is the existence of
cash''.}  See Item~\ref{i:elb}.

\item\textit{Page 52: ``One of the digital pound's principal aims is to support
payments innovation by the private sector''.}  Supporting payments innovation
requires active public engagement and debate in the process of determining
requirements and must not favour incumbent business models to the exclusion of
alternatives.  See Items~\ref{i:schumpeter} and~\ref{i:design-phase}.

\item\textit{Page 54: ``the platform model we have proposed best meets our
criteria''.}  We have not determined by consensus that the set of criteria
established by the Bank is the right set of criteria to use.  We encourage the
Bank to consider the criteria identified in this response document.

\item\textit{Page 65: ``the creation of a new sandbox for firms''.}  An
important limitation of sandboxes is that they intrinsically favour incumbents
and others with the ability to commit substantial resources to a project, in
contrast to requests for proposals for academic research funding as well as
platforms for open, public engagement.  For some reason, mention of approaches
such as these, which fundamentally enable debate, co-design, and consensus
about requirements, have been absent from the Consultation Paper.  Debate,
co-design, and consensus are instrumental for reaching a good outcome that
serves the public interest.

\item\textit{Page 77: ``Users would be likely to make digital payments using
smartphones or cards''.}  Certainly the option to use smartphones or cards
might be desirable, although cards and smartphones have serious security
weaknesses worth considering.  For example, cards as currently implemented
require users to accept the potential for misbehaviour on the part of POS
devices, since what POS devices display on their screens might not match what
they are actually doing.  (It is certainly possible to augment cards to have
better security, although this would require interaction between the user and
the card, perhaps via a display.)  In contrast, smartphones are sufficiently
powerful to avoid requiring users to trust POS devices, although their Internet
connectivity makes them susceptible to compromise via Internet-enabled software
and services, and they are known to serve the interests of third parties, such
as manufacturers or operating system distributors, in lieu of their owners.

\item\textit{Page 77: ``Not everyone has a smartphone, and some people find
them difficult to use''.}  An open architecture that allows a plethora of
non-custodial wallet implementations will provide essential support for special
CBDC use cases and for users with special accessibility needs.

\item\textit{Page 85: ``Financial inclusion means that everyone... has access
to useful and affordable financial products and services''.}  It is more
foundational, and more important, for all individuals to have access to a means
of payment than it is for all individuals to have access to credit or insurance
products.  Financial inclusion means that individuals will be able to engage
with the economy via a public payment option, without relying upon a persistent
relationship with a financial custodian.  Furthermore, arguments that access to
credit and insurance hold the answers to fundamental challenges of social
justice, such as poverty, homelessness, or statelessness, are dubious at best.

\item\textit{Page 86: ``millions of people continue to use cash across the UK,
particularly those in vulnerable groups''.}  To address financial exclusion, it
is important to ensure that a public payment option is available for all
payments.  This is not currently the case for vendors that refuse cash,
including most e-commerce vendors.  See items~\ref{i:fragmentation}
and~\ref{i:banknote}.  Note also that payment methods involving cards and
smartphones are not options for persons with special accessibility needs, such
as those with limited vision.

\item\label{i:inclusion}\textit{Page 86: ``Digital inclusion needs to be
promoted alongside financial inclusion''.}  One might believe that the fact
that not everyone has a card or a smartphone constitutes a problem that must be
solved, and that can be solved by giving everyone cards or smartphones.  This
is a dangerous misconception.  Importantly, we cannot force individuals to
establish persistent relationships with service providers, or to agree to the
terms and conditions of the services they provide.  Providing low-cost tools,
such as devices that can directly hold CBDC, might be a suitable alternative.

\item\textit{Page 88: ``Cash is expected to continue to play a role in
society... for some time''.}  The need for a means by which individual persons
can directly possess and control their own assets and engage with the economy
without fear of profiling or discrimination is foundational, and this need will
never disappear.  Currently, cash is the only general-purpose instrument
recognised by governments for use in this capacity.  Hence, the case for
digital cash is strong.

\item\textit{Page 88: ``The digital pound would not seek to replace cash''.} It
is critical that CBDC must not be designed to supplant or eliminate cash, and
CBDC should not be used to justify the expungement of cash, which for some
people and in some circumstances is intrinsically more suitable than any
digital payment option, including CBDC.  Every payment method has its own
unique set of costs and risks, and not all use cases served by cash can be
served by CBDC, however it is designed or implemented.  See Item~\ref{i:cash}.

Importantly, wallets are not accounts; see Items~\ref{i:provided}
and~\ref{i:directly}.  Apropos financial exclusion, the question should really
be whether and how service providers should be able to transfer digital pounds
into the possession and control of individuals with limited identification
documentation.  Note that limited requirements for identification does not
imply improved privacy; see Item~\ref{i:identity-assurance}.

\item\textit{Page 89: ``vulnerable groups... could be at risk when using a new
payment method''.}  The privacy risk that can lead to discrimination and
profiling is especially dangerous to vulnerable groups, such as victims of
abuse; witnesses to crime; refugees; stateless persons; persons who are members
of protected categories, such as those related to race, religion, or sexual
orientation; persons with pre-existing medical conditions; persons who have
been blacklisted as a result of a history of criminal convictions or financial
default; persons in abusive personal, domestic, or business relationships; and
so on.

\item\textit{Page 89: ``not necessarily the only way to tackle financial
exclusion''.}  Another way to tackle financial exclusion might be to provide
assurance that cash will continue to function as a means of engaging with the
economy.  This can be achieved by establishing law or regulatory policy to
require its acceptance by brick-and-mortar vendors at the point of sale, as
well as by e-commerce vendors, many of which currently do not have a means to
accept cash as a means of payment.  See Item~\ref{i:compel}.

\item\textit{Page 90: ``sceptical about the current need for a retail CBDC in
the UK, which they considered to already have an efficient payments system''.}
The efficiency of payments in the UK is limited by the lack of a public payment
option suited to the increasingly digital economy, as well as by the pernicious
refusal of cash by vendors, which causes some transactions that would otherwise
occur to not occur, an direct example of inefficiency.

\setcounter{enumTemp}{\theenumi}
\end{enumerate}

\subsection{Privacy}

\begin{enumerate}
\setcounter{enumi}{\theenumTemp}

\item\label{i:like-bank-accounts}\textit{Page 11: ``Privacy protected like
cards and bank accounts''.}  But cards and bank accounts are not private at
all.  Data consumers such as providers of credit and insurance, as well as law
enforcement organisations, routinely have access to transaction data, including
the identities of payers and payees.  In addition, data breaches by criminal
organisations and state actors are not uncommon.  So, a goal to deliver a
comparable level of privacy to cards and bank accounts is a goal to not deliver
privacy.

\item\textit{Page 11: ``but not anonymous''.} The Consultation Paper fails to
distinguish between transactions that are entirely beyond the view of
regulators and transactions that are broadly visible to regulators except that
some information about the transactions are not visible.  The Consultation Paper
also fails to distinguish between ``fully anonymous'' transactions in which
authorities know none of the details about the transacting parties, and
``semi-anonymous'' transactions in which only one party (for example, the
payer) is anonymous.

\item\label{i:exceptional}\textit{Page 11: ``BOE/government would not see any
personal data''.} With the design proposed in the Consultation Paper, this claim
is not true, since the linkages among transactions are visible, and collections
of transactions known to belong to the same person are indeed PII.  Also, if
profile data (combinations of account-holder identifiers collected via KYC
procedures, ``alternative data'' such as what can be gleaned form smartphone
apps and websites, transaction metadata such as timing and location,
transaction linkages via pseudonymous identifiers, and so on), are
discoverable, then there is no way for a user to verify that they have not been
discovered and used, for example in the controversial practice of parallel
construction~\cite{hrw-parallel}.

\item\textit{Page 11: ``to safeguard their privacy''.}  But parties with access
to PII can collude to determine the owner, so privacy is not assured.  In
particular, with the design described in the Consultation Paper, the central bank
still sees assets through its management of the core ledger and can collude
with other parties to help them determine how individual consumers spend their
money, as well as other private information such as their movement, social
networks, financial circumstances, and more.  The likelihood of surveillance
introduces a chilling effect to their free engagement with the
economy~\cite{armer1968,armer1975}.

\item\label{i:majority}\textit{Page 12: ``Digital payments account for the
majority of transactions today.''}  The lack of privacy in digital payments
today is an important motivation for deploying CBDC that is private by design,
and it does not constitute a valid argument for why it would be acceptable for
CBDC to have a similar level of privacy to existing digital payment mechanisms.
The fact that digital payments constitute an increasing share of all payments
indicates that the right to privacy is being eroded by the migration of retail
payments to the digital economy.  CBDC can be part of the solution, but only if
it adopts privacy by design for consumers.

\item\label{i:identify}\textit{Page 12: ``the digital pound would not be
anonymous because the ability to identify and verify users is needed to prevent
financial crime''.} If CBDC is to be used as a digital form of cash that is
obtained via regulated institutions with AML/KYC procedures, then it should
have similar privacy features for consumers who obtain physical cash from
regulated financial institutions with AML/KYC procedures and subsequently spend
the cash without linking their identities to the transactions wherein they
spend the cash.  The ability to identify and verify users when they receive
CBDC does not imply the ability to link users to their payments.  A system that
allows users to make payments without the risk of profiling does not imply that
users who receive cash would not be identified and verified when they receive
CBDC, either by making a withdrawal of CBDC, or by receiving a payment in CBDC.

\item\textit{Page 12: ``the digital pound would be at least as private as
current forms of digital money, such as bank accounts''.}  See
Item~\ref{i:like-bank-accounts}.

\item\textit{Page 12: ``except for law enforcement agencies under limited
circumstances prescribed in law''.}  This implies an exceptional access
mechanism, otherwise known as a ``backdoor'', of the sort that is considered
dangerous and untenable by the security
community~\cite{abelson1997,abelson2015,benaloh2018}.  Procedures for gaining
exceptional access to private data would not be established if there were no
expectation that they would be used from time to time.  Exceptional access
mechanisms introduce the likelihood that data will be abused without
consequence, not only by legitimate authorities, but also by other authorities
who request access, unscrupulous insiders who abuse their position, businesses
that operate the infrastructure, criminal attackers or foreign state actors who
compromise the infrastructure, and more.  See Item~\ref{i:exceptional}.

\item\label{i:same-basis}\textit{Page 12: ``on the same basis as currently with
other digital payments and bank accounts more generally''.}  Consumers cannot
trust what they cannot verify.  With cash, privacy for consumers is
self-evident.  Currently, what consumers have with digital payments is better
described as ``privacy by promise'': there is no option for consumers but to
accept that service providers will have a great preponderance of their personal
transaction data in their databases.  Considering that many retail vendors in
the UK, including e-commerce vendors as well as brick-and-mortar vendors at the
point of sale, refuse cash, this is not so much a matter of trust as it is a
matter of coercion; see Item~\ref{i:like-bank-accounts}.  Also, the argument
that it must be suitable to assume a similar privacy model to current digital
payments because digital payments are commonplace implicitly relies upon the
assumption that a one-size-fits-all approach to privacy is appropriate, which
it is not; see Item~\ref{i:majority}.  The lack of privacy by design is
particularly harmful to vulnerable consumers, such as those in precarious
domestic relationships, victims of abuse, witnesses to crime, refugees, and
others.  Privacy is a public good~\cite{fairfield2015}, and intrinsic respect
for privacy, as opposed to a demand to accept panopticon-like surveillance,
should be a requirement for public infrastructure in general.  Although it can
be selectively taken away from a system that is private by design, privacy
cannot be added to a system that is not private by design.

\item\textit{Page 13: ``providing the same privacy as most of the money we
use''.}  It might be true that most retail transactions are digital, but not
everyone uses digital payments for most of their retail transactions.  Persons
who prefer or depend upon physical cash today risk being coerced into accepting
the privacy model underpinning current digital payments, which is inadequate.
See Items~\ref{i:like-bank-accounts}, \ref{i:majority}, and~\ref{i:same-basis}.

\item\textit{Page 14: ``The Bank and HM Treasury would welcome views on this
approach to privacy''.}  Privacy by promise is not real privacy, and the
privacy model for current digital payments used by retail consumers is not
appropriate for CBDC.  Consumers deserve affirmation of their right to make
payments without being associated with every payment they make, a right that
has existed for thousands of years and that is undermined by the migration to
the current set of digital payments.  Consumers in particular deserve a payment
system design that does not demand that consumers accept paternalistic
collection of data that offer insight into their individual habits, locations,
decisions, and modes of life.  As Paul Armer warned a half-century ago, ``If
you wanted to build an unobtrusive system for surveillance, you couldn't do
much better than an EFTS [electronic funds transfer system]''~\cite{armer1975}.

\item\label{i:privacy-competition}\textit{Page 46: ``data advantages''.}  The
benefit of surveillance to well-resourced firms, and deleterious effect on
competition, can be mitigated through privacy by design.

\item\textit{Page 51: ``Our proposed model... safeguards data protection and
privacy''.}  The privacy model proposed in the Consultation Paper is privacy by
promise, not privacy by design.  See Items~\ref{i:exceptional}
and~\ref{i:same-basis}.

\item\textit{Page 53: ``The Bank... would not know the identity of the
payer''.}  The privacy risk is not mostly about the Bank, but parties other
than the Bank, such as authorities who compel access, criminals who seize
access, and businesses who operate the system, using exceptional access
mechanisms to determine, access profiling information.  See
Item~\ref{i:exceptional}.

\item\textit{Page 57: ``PIPs might use transaction data to improve existing
operations or to offer new customer-facing services''.}  Functions involving
the collection and use of transaction data are similar to bank accounts, with
all of the same implications for privacy.  See
Items~\ref{i:like-bank-accounts}, \ref{i:same-basis},
and~\ref{i:personalisation}.

\item\textit{Page 57: ``Cross-subsidisation [can potentially lead to]
dependence on associated non-payments activities for the provision of critical
digital pound services''.}  Cross-subsidy is particularly risky given the
opportunity to collect data revenues from credit, insurance, and blocklisting
services, or from data brokers more generally.  Data revenues may be a
significant contributor to revenue for businesses that engage in narrow
banking.  See Item~\ref{i:narrow}.

\item\label{i:cajoling}\textit{Page 67: ``Transparency and clear understanding
of the rights and tools around personal data will promote good data use''.}  No
quantity of cajoling or process transparency can convince the public that data
collectors will keep their promises to keep data secure and to not misuse data,
to say nothing of the fact that they are not promising to not share data with
authorities who also might fail to keep data secure and not misuse data.
``Privacy by promise'' is no substitute for privacy by design.  No one believes
that government authorities would not have access to the transaction data of
individual consumers if they were to ask in the right way, just as no one
believes that the systems that store and process data would be impervious to
malicious attackers, accidental or deliberate data breaches, signals
intelligence capabilities of foreign state actors, and the quotidian business
interests of the private-sector businesses that will ultimately own and operate
much of this infrastructure.  To expect otherwise is fanciful at best, and to
employ coercion to compel consumers to share information about how they spend
their money undermines their fundamental right to use the money as they wish,
which is to say that it would undermine their right to ownership of money in
general.

\item\textit{Page 68: ``Digital transactions account for the majority of
transactions in the UK today''.}  The primary motivation for retail CBDC is the
decline of the use and acceptance of cash, which implies the lack of a public
payment option in practice.  Today, consumers in the UK are rapidly losing the
option to use cash for their ordinary purchases.  To argue that consumers must
simply accept the same lack of privacy that applies to card payments and bank
transfers is tantamount to arguing that the level of privacy offered by modern
retail payment systems does not create negative externalities, which it does,
and is good enough for everyone in every circumstance, which it is not.  See
Item~\ref{i:majority}.

\item\textit{Page 68: ``Personal data from bank account transactions is used
and stored by firms to comply with legal and regulatory data capture
requirements''.}  The privacy threat of personal data capture results from the
assumption that consumers cannot have custody of their own money.  Legal and
regulatory data capture requirements that associate consumers with transactions
apply to transfers from custodial accounts.  Such requirements apply to the
custodians and do not apply that the identity of a consumer must be associated
with all of the economic transactions that the consumer makes.  Consider that
with cash transactions, consumers have the right to engage with the economy
without revealing their identities when they do.  This right should be
affirmed, not curtailed, by lawmakers and regulators.  Although the CBDC
architecture does not permit consumers to possess and control their own CBDC
assets, alternative architectures that do so should be considered instead.
Architectures that allow users to possess and control their own assets would
enable users to transact their own assets privately without violating rules
related to the disclosure obligations of regulated custodians.

\item\textit{Page 68: ``firms also have to comply with UK data protection
laws''.}  Law cannot prevent data breaches, does not stop government
authorities from misusing the data to which they have legal access, and does
not stop insiders from accessing the data to which they have technical access.
Regulations such as GDPR have not been effective in preventing such misuses of
data~\cite{staschen2022}, and users cannot trust what they cannot verify.  See
Items~\ref{i:like-bank-accounts} and~\ref{i:exceptional}.

\item\textit{Page 69: ``the UK GDPR requires that firms must be satisfied that
sharing personal data with a law enforcement authority is lawful''.}
Regulations that compel regulated businesses to cooperate with authorities do
not necessarily mean that consumers must be subject to profiling.  It is not
that businesses that facilitate transactions should be uncooperative, but that
the design of the system should not generate data that makes it possible to
connect consumers to their payments.  In particular, if consumers are able to
possess and control their own assets, then it will be possible for them to
transact without involving custodians who know their identities.  See
Items~\ref{i:provided} and~\ref{i:directly}.

\item\textit{Page 69: ``at least as private as current forms of digital
money''.}  Current forms of digital money are not private, and a central
motivation for the exploration of CBDC is the disappearance and refusal of
cash, which is more private.  See Items~\ref{i:majority}
and~\ref{i:same-basis}.

\item\textit{Page 69: ``exert greater user control of personal data''.}  The
only way for users to be in control of their transaction data is to not
associate their identities with the transactions in the first instance, which
also implies that users must have a way to avoid associating their transactions
with each other.

\item\textit{Page 70: ``except for law enforcement agencies under limited
circumstances''.}  This describes an ``exceptional access mechanism''; see
Item~\ref{i:exceptional}.

\item\textit{Page 70: ``the ability to identify and verify users is needed to
prevent financial crime''.}  Identifying and verifying users does not imply
associating payers with their transactions.  See Item~\ref{i:identify}.

\item\textit{Page 70: ``greater benefits from sharing their personal data''.}
The potential for consumers to benefit do not justify compulsory revelation of
all spending habits and patterns, as is often the case with digital payments
today, as payment service providers have business models that depend upon the
potential, for themselves or for customers of their data, to analyse the
behaviour of their customers.  Also, there is no evidence offered to support
the proposition that users actually do benefit from sharing their data; the
proposition may be valid but unsound.

\item\textit{Page 74: ``There is public appetite for trading personal
information for access to products and services''.}  The argument that personal
information can be monetised for the benefit of data subjects is valid but not
sound.  An individual's data is generally more relevant, and thus more
valuable, to the data subject than to other recipients.  Additionally,
effective data brokers generally amass such large datasets that the marginal
value of additional data related to a single individual is usually close to
zero (submodularity).  Finally, encouraging people to reduce their privacy by
revealing their data creates negative externalities by reducing the size of the
set of parties who have not chosen to do the same, thus reducing their privacy
as well.

\item\textit{Page 87: ``understanding and trust among the public is crucial''.}
Trust is not achieved by requiring people to provide their identities in every
transaction, or to require them to store their assets with custodians as a
condition of being able to make a payment.  See Item~\ref{i:cajoling}.

\setcounter{enumTemp}{\theenumi}
\end{enumerate}

\subsection{Custody}

\begin{enumerate}
\setcounter{enumi}{\theenumTemp}

\item\label{i:provided}\textit{Page 11: ``Wallets to hold digital pounds
offered by the private sector''.}  If users are forced to engage with
custodians in the course of owning digital assets, then they do not really have
the degree of possession or control suggested by the use of the term
``wallet''.  It should be possible for owners of CBDC to use open-source tools,
not just services, that allow them to possess and control CBDC directly.

\item\label{i:directly}\textit{Page 11: ``Accessed by users through smartphones
or cards''.} The word ``accessed'' suggests that the tokens must be stored by
some custodian, which controls access; actually, it must be possible for tokens
to be stored directly on the devices, not merely accessed.  It may be that
users will be able to access CBDC using smartphones or cards, although it must
be possible for users to use unregistered devices as well.  In addition,
smartphones and cards usually important security weaknesses.  Smartphones
usually have a large attack surface and remote-access vulnerabilities that
result from being connected to the Internet, while cards usually lack a display
or input mechanisms, meaning that users are required to trust third-party
hardware such as ATMs and point of sale devices.

\item\textit{Page 11: ``Limited amount per user''.}  Limiting the amount of
CBDC that a user can hold implies that either a representation of the amount of
CBDC that a user has is effectively stored in a ledger somewhere, making it
functionally not different from an account, or the wallets that hold CBDC are
not really under the control of their owners, since they must enforce rules
contrary to the desires of their owners.  Both scenarios imply that users do
not really have control of their money.

\item\textit{Page 13: ``a digital pound... would be available to non-UK
residents too''.}  Visitors require a way to access the UK digital economy as
well, and it is appropriate to expect that they would own digital pounds just
as visitors own physical pounds.  However, it is less obvious that such
visitors would need to establish persistent relationships with UK banks, as is
an implicit requirement of the design described in the Consultation Paper.  A
better approach would be to allow owners of CBDC to hold the assets directly,
as they do with physical cash.  Correspondingly, it would be reasonable for
visitors who exchange foreign currency (or CBDC) for UK CBDC to avoid the
requirement to provide personal information for AML/KYC purposes, provided that
they exchange the assets in-person and that the service provider that offers
the exchange strictly enforces limits, as is done with travel money exchanges
(bureaux de change).

\item\textit{Page 13: ``personal details would be known to their private sector
wallet provider''.}  Consumers must have a way to directly possess and control
the CBDC directly, using tools rather than services.  Otherwise, the money is
not really theirs but rather something that they are only conditionally
privileged to use, subject to the conditions and decisions of their service
providers.  See Item~\ref{i:provided}.

\item\label{i:receipt}\textit{Page 14: ``Unlike cash, the amount of digital
pounds an individual or business could hold would be subject to some
restrictions... to ensure a smooth introduction without unintended consequences
for monetary or financial stability''.}  The goal of avoiding unintended
consequences can be realised by imposing restrictions on withdrawals of CBDC
from banks, just as it can be realised by imposing restrictions on withdrawals
of cash from banks.  Applying a limit to withdrawals rather than holdings is
more direct and appropriate mechanism to safeguard against bank runs, since it
is really the withdrawals by consumers and businesses, not the total holdings
by consumers and businesses, that potentially introduce strain on the balance
sheets of banks.  The choice to impose the restriction on holdings rather than
withdrawals is therefore not justified.

We note also that limits on receipt or withdrawals can implicitly be used to
implement limits on total holdings.  For example, a regulator, or
private-sector bank through its risk management practice, might decide to limit
aggregate CBDC receipt or withdrawals to some amount each day, or each month,
and the issuer might establish an expiration date for a particular CBDC
vintage~\cite{goodell2021a}.  These mechanisms can be used to mitigate the risk
of bank runs without managing the holdings of individual consumers.

Restrictions on the amount of CBDC an individual or business can receive or
withdraw are different from restrictions on the amount of CBDC an individual or
business can hold.  In particular, restrictions on receipt or withdrawals can
be imposed by banks or other service providers involved in distributing and
redeeming CBDC, whereas restrictions on total holdings imply either the
implementation of an account-like structure for the owner, which would be
managed by the central bank or a private-sector business, or the reliance upon
``wallets'' to enforce rules about what they can hold, for example by using a
wallet registration system or trusted hardware, against the interests of the
owners of the CBDC.  This design choice is somewhat bizarre, since it suggests
a deliberate protection of the current business models of incumbent providers
of services that provide custody.

\item\textit{Page 14: ``It would be for a further decision, in the light of
experience, whether those restrictions should be made permanent''.}  It is
certainly prudent to make a decision following some introductory period,
although we imagine that there should be some permanent restrictions,
accounting for factors such as inflation and distribution of wealth, although
the restrictions should be like cash: on receipt and withdrawal of CBDC, not
aggregate holdings.  Restrictions on receipt and withdrawal of CBDC are
sufficient to protect against exigent threats such as bank runs.

We also note that as retail payments migrate from cash to the digital economy,
the right to hold an unlimited amount of physical cash might be undermined, as
physical cash becomes less useful.  So, we will need a way to preserve this
right in the digital economy, by explicitly not imposing a limit on total
holdings.

\item\textit{Page 19: ``DeFi applications... enable users to buy, swap, sell
and settle crypto products wihout reliance on central intermediaries or
institutions.''}  Arguably, the purpose of DeFi is to enable the direct
exchange of value without requiring transacting parties to make use of
custodians or accounts.  This concept is akin to digital cash and is worthy of
support.  Note that it is possible to avoid imposing a requirement for
consumers to invoke custodians or accounts in digital retail transactions
without decentralised issuance and without undermining regulatory compliance
requirements.

\item\textit{Page 25: ``financially risk-free money widely accepted for
transactions in the UK will be available in both stress and normal times''.}
Centralised approaches wherein users are not directly in possession and control
of their assets are vulnerable to shocks wherein centralised system operators
are damaged or disabled, or when communication links between transacting
parties and centralised system operators are severed.

\item\textit{Page 28: ``giving households the security that they could exit the
banking sector or private payment platforms to a digital, financially risk-free
asset''.}  Another critical aspect of this security is knowledge that users
cannot be blocked by asset custodians.  Consider the case of protesting
Canadian lorry drivers~\cite{woolf2022}, whose bank accounts were blocked by
authorities.  The proposed architecture does not address this risk, because of
the role of ``wallet'' service providers as intermediaries.

\item\textit{Page 29: Contactless card payments, mobile payment apps, payment
facilitators, Open Banking.}  The mechanisms listed all facilitate and promote
the use of custodial accounts as a means of payment rather than cash.  Although
Open Banking was promoted as a way to disintermediate retail payments, the use
of card payments have only increased since the introduction of Open Banking.
Although Open Banking might have reduced costs of card operations or the total
cost of interbank transactions, interchange fees have remained high,
demonstrating the weakness of the supply-side argument that transacting parties
(consumers and vendors) would benefit.

\item\textit{Page 32: ``smart contracts, which carry out specific actions based
on pre-defined terms and conditions''.}  In principle, CBDC can interoperate
with a rich framework for smart contracts.  However, smart contracts that serve
to bind a user to an agreement with a third party, or to impose other
conditions on the use of the assets in question, necessarily require the user
to not be in possession or control of the assets.  See Items~\ref{i:provided}
and~\ref{i:directly}.  Since the ability for consumers to directly possess and
control digital assets is an essential design feature, execution of smart
contracts that enforce rules or impose conditions or restrictions are perhaps
best handled by the use of an escrow agent.

\item\label{i:narrow}\textit{Page 34: ``includes the potential for... narrow
banks''.} Deposits at narrow banks are generally expected to be backed by
liquid sovereign debt rather than risky securities, and the primary function of
such banks is to manage deposits and facilitate payments rather than to engage
in credit creation.  However, given that financial intermediation is the
traditional role of commercial banks, it is worth considering the implications
of narrow banking as a business model.  It is not obvious that consumer-facing
banks should have a role in retail transactions, for which there are important
negative externalities.  When consumers use cash, their banks are not involved;
the rise of narrow banking suggests an increasing role of the custodial
accounts of consumers in retail commerce, with a concomitant deleterious impact
on the rights of consumers to privacy, possession, and control (see
Item~\ref{i:cash-role}) and a concomitant deleterious impact on financial
inclusion and payments resilience (see Item~\ref{i:resilience}).

\item\textit{Page 34: ``whether new forms of private digital money display
adequate interoperability''.}  New forms of private digital money often drive
fragmentation.  See Item~\ref{i:fragmentation}.

\item\textit{Page 36: ``offline payments... could be valuable in remote
areas''.}  If, by the term ``offline payment'', what is meant is a payment
wherein neither party has a data connection at all, then it is natural to
consider solutions that involve trusted hardware.  However, although trusted
hardware may be a valid option for transacting parties who freely choose to
trust the hardware manufacturer, in such cases, the hardware manufacturer
becomes a \textit{de facto} custodian, despite the fact that the owner of the
assets might have possession of the device holding the assets.  Establishing
CBDC that \textit{requires} certified hardware, including hardware with
``secure enclaves'', introduces many problems and negative
externalities~\cite{goodell2022a}.

Alternatively, if ``offline payment'' means only that the transacting parties
do not have access to the Internet, then it is technically possible for the
transacting parties to trust a local third party, rather than a system-wide
``core ledger'', to memorialise some record of the
transaction~\cite{goodell2022}.  However, this requires some measure of
decentralisation, whereas the system proposed in the Consultation Paper seems to
assume that the effect of all transactions must be recorded in the core ledger.

Finally, it is worth considering whether it is actually necessary for a CBDC to
provide support for offline payments of either type.  The payment technologies
that have supplanted cash in recent years usually rely upon fully functioning
network connections.  Today, there are many different payment mechanisms,
including cash, and each mechanism carries its own set of costs and benefits.
Even in the long-term, it is unreasonable to expect that CBDC will outperform
or replace every payment solution in all circumstances, including cash, which
performs well as a robust offline payment solution.

\item\textit{Page 40: ``Limits on holdings of the digital pound... would
constrain the extent of outflows from bank deposits''.}  The statement is true,
although limits on holdings is not the only way or the best way to constrain
the extent of outflows from bank deposits.  The extent of outflows from bank
deposits can be constrained directly, by limiting the rate at which individuals
can withdraw or receive CBDC.  Most importantly, by limiting receipt and
withdrawals rather than limiting holdings, users can possess and control their
own wallets, which is a necessary prerequisite to possessing and controlling
their own money.  See Item~\ref{i:receipt}.

\item\textit{Page 40: ``there would be uncertainty about... banks' ability to
replace lost retail funding with wholesale funding''.}  There are two issues
here: the effect of having current accounts that are smaller by the amount of
cash that individual consumers would have carried with them prior to the
burgeoning popularity of card payments over the past decade, and the effect of
lost revenue associated with transaction fees and data harvesting associated
with payment processing activities.  The former can be expected to be modest;
the latter is about banks relying upon an activity other than banking as a
critical source of revenue.  See Item~\ref{i:switch}.

\item\textit{Page 46: ``new forms of digital money... could cause commercial
banks to lose some of their retail deposits''.}  See Items~\ref{i:switch}
and~\ref{i:deposits}.

\item\textit{Page 47: ``commercial incentives... could favour the creation of
`walled gardens' with low interoperability''.}  Walled gardens, including not
only card payment networks but also their principal competitors, such as mobile
payment apps, also undermine consumer privacy and control.  See
Item~\ref{i:privacy-competition}.

\item\label{i:safes}\textit{Page 49: ``Relative to cash, the digital pound
would have negligible storage costs''.}  Although it might seem that because a
simple USB stick can store an arbitrarily large quantity of digital assets or
the keys that control them, this statement is not justified.  For example, the
cost of a security safe is generally a function of the value, and not the
physical size, of what is stored inside~\cite{goodell2021a}.  The operational
cost of managing backups and securing digital devices against theft is not
trivial, and not obviously less than managing and protecting cash.

\item\textit{Page 49: ``an unremunerated digital pound could make it more
difficult or banks to charge negative deposit rates without losing deposits''.}
See Item~\ref{i:elb}.

\item\textit{Page 51: ``Our proposed model... promotes accessibility''.}  A
better way to promote accessibility is to allow a means for owners of CBDC to
possess and control it directly, rather than to require it to be ``provided''.
See Item~\ref{i:provided}.

\item\label{i:never-in-possession}\textit{Page 53: ``The private sector would
never be in possession of end users' digital pound funds''.}  With assets
represented directly on the core ledger, managed and discoverable by core
ledger operators or those with the power to compromise or compel access to the
information on the ledger, end users would never be in possession of their own
CBDC funds, either.

\item\textit{Page 55: ``PIPs, and the wallets they provide, would never be in
possession of end users' digital pound funds''.}  But PIPs can dictate terms in
which wallets operate.  See Item~\ref{i:never-in-possession}.

\item\textit{Page 61: (delegated model) ``the PIP... has a record of a user's
holdings of digital pounds''.}  A user's holdings must not be subject to
surveillance of this kind.  With the ``delegated model'' described in the
Consultation Paper, the user must have an account with the PIP, which
intermediates a consumer's transactions, not unlike electronic payments using
traditional bank deposits.  See Item~\ref{i:account}.  Unfortunately, this
architecture, like the primary architecture proposed in the Consultation Paper,
implies that users are not really in possession or control of their own assets,
and they cannot really transact privately.  A better approach would allow PIPs
to monitor withdrawals of CBDC into non-custodial wallets in the possession and
under the control of a consumer.  Once in the non-custodial wallet, the PIP no
longer has visibility of the assets and does not learn when they are spent.
Thus, the PIP would have record of deposits and withdrawals but would not have
knowledge of aggregate holdings, because the PIP would not know whether or when
assets are spent.

\item\textit{Page 61: (delegated model) ``greater technical and operational
requirements on PIPs''.}  In principle, there is a design choice concerning
whether a PIP would maintain its own ledger or not.  Having PIPs maintain their
own ledgers, as in the ``distributed model'', would improve scalability, and is
more consistent with models based on self-validating tokens and oblvious
transfers, wherein neither the ledger trusted by the issuer nor the ledger
maintained by a PIP actually store representations of tokens on a ledger at
all.

\item\label{i:bearer-1}\textit{Page 61: (bearer instrument model) ``no trusted
intermediary...  could give rise to `double spend risk'''.}  The ``bearer
instrument model'' described in the Consultation Paper is a straw man.  It is
theoretically impossible to have fair exchange without a trusted third
party~\cite{pagnia1999}, and the ``double spend risk'' referenced in the
Consultation Paper is not inherent to the requirement for users to possess and
control their own assets.  Specifically, an asset can reference (implicitly or
explicitly) a specific trusted intermediary or network that must be used to
facilitate its next transaction.  This method allows users to have fair
exchange whilst possessing and controlling their own assets.

\item\label{i:bearer-2}\textit{Page 61: (bearer instrument model) ``a bearer
instrument approach... would lead to completely anonymous payments''.}  As
noted in Item~\ref{i:bearer-1}, the ``bearer instrument model'' described in
the Consultation Paper is a straw man; there are other models for bearer
instruments that are not considered by the Consultation Paper.  The Consultation
Paper seems to assume that if a consumer is able to have possession and control
of a CBDC asset, then it must be possible:

\begin{enumerate}

\item to have peer-to-peer transactions without any involvement of third
parties;

\item to have a transaction without reporting the transaction to regulators;
and

\item for both counterparties in a transaction to be anonymous.

\end{enumerate}

In fact, none of these assumptions are true.  It has been shown that an
architecture with self-validating tokens and oblivious transfers can be built
to allow consumers to directly possess the tokens that they use, while also
using Chaumian blind signatures~\cite{chaum1982,chaum2021} to make the sender
anonymous.  In a system of this type, which is not considered in the disussion
paper, payers are anonymous and payees are not.  The asset is absolutely a
``bearer instrument'' in the sense that the payer holds a fungible asset
directly, does not provide identity information during transctions, and is not
subject to the rules of a custodian.

However, regulators could specify rules to ensure that payments are not
completely anonymous, for example, by requiring identifying information of the
recipient to be embedded into the asset with each transaction and requiring
that tokens can be transacted only once before being converted into bank
deposit or exchanged by a regulated intermediary for a new asset.

\item\textit{Page 61: (bearer instrument model) ``There is additional
complexity for conducting transactions between two individuals over
distance''.}  The assumption that having bearer instruments implies conducting
transactions between individual consumers is false.  Transactions can be
facilitated by third parties without requiring those third parties to be
custodians.  The Consultation Paper should not have dismissed direct possession
and control by end users outright, simply because of the shortcomings of a
specific payment architecture described as using ``bearer instruments''.  See
Item~\ref{i:bearer-2}.

\item\textit{Page 64: ``based on open standards (including ISO 20022 for
messages)''.}  When ISO 20022 was introduced, most consumer retail payments in
the UK still used cash, and it was assumed that all legitimate electronic
transactions would have custodians on both sides of the transaction,.  However,
digital currency introduces the prospect of electronic bearer instruments and
the possibility that digital tokens could be held outside accounts.  ISO 20022,
therefore, is not built for digital currency and would require refreshing if it
is to be used for that purpose..

\item\label{i:wallet-verification}\textit{Page 72: ``identity verification
would be required when opening a digital pound wallet''.}  The architecture
described in this Consultation Paper seems to conflate ``wallets'' with accounts.
Wallets are tools to hold digital assets; see Items~\ref{i:provided}
and~\ref{i:directly}.  A better way to enforce identity verification is to
require identification at the point of withdrawal or receipt of CBDC from a
regulated bank or service provider into a digital wallet, as is done with bank
withdrawals in which cash is received.  This is not the same as requiring the
wallet itself to be identified.

\item\textit{Page 72: ``PIPs would require identity information of wallet
account holders''.}  Wallets are not the same as accounts; consumers should
have the option to hold assets directly.  See Items~\ref{i:provided}
and~\ref{i:directly}.  Having PIPs collect information about their customers
does not imply involving PIPs in the transactions wherein consumers make
payments.  See Item~\ref{i:wallet-verification}.

\item\textit{Page 72: ``the UK AML and CFT Regime dictates that additional
information about the payer must be collected for large-value transactions made
in cash''.}  However, AML/CFT regulation does not require payer information for
small-value transactions.  To preserve this right, it is necessary for
consumers to have a way to make small-value transactions without associating
payer information with the transaction, as it is for payments with cash.
Unlike with cash, a CBDC system could ensure that the recipient is fully
monitored, thus offering a greater degree of compliance enforcement on the
transaction without compromising the privacy of the consumer.  A risk-based
approach would entail requiring recipients of CBDC to explicitly collect payer
information under some circumstances, including but not limited to high-value
payments.

\item\textit{Page 72: ``The digital pound would have lower frictions than
physical cash, so carries higher risks of abetting crime''.}  This statement is
false.  Frictions depend upon the design of the system.  For example, if it is
possible to monitor all of the recipients without revealing the identities of
the payers, it is possible to both see all transactions and impose frictions in
a way that is not possible with cash.  Also, transactions can be monitored for
structuring via a risk-based approach based upon analysing receipts and
withdrawals of CBDC.  See Item~\ref{i:safes}.

\item\label{i:instability}\textit{Page 80: ``risks largely stem from any large
and rapid outflows into digital pounds... [a] limit on individual holdings
would be intended to manage those risks by constraining the degree to which
deposits could flow out of the banking system''.}  The risk of financial
instability resulting from large and rapid outflows from bank deposits is a
reasonable concern, although a limit on individual holdings is neither
necessary nor sufficient to address that concern.  A limit on holdings is not
sufficient, because a limit on holdings does not mean that it will be possible
to prevent many users from making maximum-sized withdrawals during a crisis.  A
limit on holdings is not necessary, because there are other, better ways to
limit withdrawals during a crisis, including limiting the size or rate of
withdrawals directly.  Such approaches can also be introduced rapidly during
times of crisis, as they have been with cash~\cite{ap2018}.  See
Item~\ref{i:atm-limits}.

\item\label{i:salary}\textit{Page 80: ``users may want to use their digital
pound wallet to receive their salary''.}  Although this may be possible, it
might not be appropriate in most cases.  Employees and service providers
commonly want third-party proof that employers have paid them, and this
function is usually performed by the provider of an account, not a tool for
holding assets.  In addition, consumers might also prefer receiving money into
a bank account because they will have an opportunity to earn interest, receive
consolidated reports for tax compliance purposes, or conveniently invest in
securities.  Accounts offer many benefits, but proper wallets are not accounts.
The Consultation Paper seems to implicitly assume that CBDC wallets would
substitute for bank accounts.  CBDC wallets are not a substitute for bank
accounts, which serve a critical role in financial intermediation, nor are they
a substitute for cash, which is intrinsically more effective than digital
payment mechanisms in certain use cases and situations.  See
Items~\ref{i:provided} and~\ref{i:directly}.

\item\textit{Page 80: ``We seek feedback on the proposed holding limit''.}
There should be limits on receipt and withdrawal of CBDC, not holding limits.
Please see the response to Question 7 in Section~\ref{s:consultation}.

\item\textit{Page 82: ``Limits would be in place at least during transition''.}
Any limits should be imposed on the receipt or withdrawal of CBDC, rather than
upon aggregate CBDC holdings.  The system must not treat CBDC wallets as a kind
of account, but as a digital asset that can be possessed and controlled in a
manner similar to how a physical assets can be possessed and controlled.  See
Items~\ref{i:provided}, \ref{i:directly}, and~\ref{i:instability}.

\item\textit{Page 83: ``How many digital pounds should corporates be able to
hold''.}  Corporate persons are different from human persons, and do not have
human rights.  Nevertheless, compared to limits on \textit{withdrawals}, limits
on \textit{holdings} are inappropriate and less well-suited to the problem of
addressing financial instability.  Also, it might be reasonable to assume that
different corporations should have different limits, which can be imposed at a
specific level for each business by requiring enforcement by custodians and
money services businesses that transfer money to the business in question.  See
Item~\ref{i:instability}.

\item\textit{Page 83: ``holdings above the level of the limit might be
automatically `swept' into a nominated bank account''.}  The Consultation Paper
assumes that a CBDC wallet is essentially a kind of account, with actual
custody of the assets under the control of a party other than the beneficial
owner.  This is problematic for several reasons; see Items~\ref{i:provided}
and~\ref{i:directly}.  In this case, the Consultation Paper also assumes that
CBDC wallets must be online with full connectivity to the designated bank
account in question.

Nevertheless, non-custodial CBDC wallets under the direct possession and
control of users can still be used in conjunction with accounts that enforce
certain rules, such as a limit on the quantity of CBDC withdrawn net of the
quantity of CBDC deposited.  The wallets can function as practical tools that
allow their users to manage their acceptance of CBDC, their deposits of CBDC,
and their withdrawals of CBDC in anticipation of enforcement of the rules by
the accounts with which they interact.

\item\textit{Page 84: ``distinguishing which type of business should or should
not have access to the digital pound''.}  It is certainly possible to prevent
certain businesses from withdrawing or accepting digital pounds.  It is also
possible for a regulator to specify the conditions in which a recipient of CBDC
will be able to redeem it.  For example, a regulator may stipulate that a
consumer must embed the identity of the recipient or even a valid bank account
of the recipient, into some or substantially all transactions.  A regulator
might also specify that the recipient must deposit the CBDC into a bank account
within a certain period of time, or even that the recipient must receive CBDC
directly into its nominated bank account, as a condition of business.

\item\textit{Page 87: ``for those without the internet or smartphones, offline
capabilities and other solutions are being explored''.}  The set of
general-purpose solutions under consideration should include those under
investigation by the Future Infrastructure for Retail Remittances (FIRE)
Project, which is being led by University College London and the University of
Edinburgh.

\item\textit{Page 87: ``need to be designed to help people access services''.}
CBDC wallets are not the same as basic bank accounts and should not be
evaluated in such terms.  Access points and means of identification should be
subject to regulation and managed by banks and other regulated money services
businesses, and they should be designed to help people receive money in the
absence of a persistent relationship with a bank or other asset custodian.  See
Items~\ref{i:provided} and~\ref{i:directly}.

\item\textit{Page 89: ``we discussed the possibility of civil society groups
becoming wallet providers''.}  It must not be a requirement that wallets would
be ``provided'' by some custodian or service provider; see
Items~\ref{i:provided} and~\ref{i:directly}.  The main issue with respect to
service providers in the context of civil society is the matter of cajoling
people to accept the authority of a custodian or other gatekeeper as a
prerequisite for engaging with the economy.  Such custodians and gatekeepers
are not necessary for the use of cash, since it is directly possessed and
controlled.  Provided that there is a means for individual persons to possess
and control CBDC directly, then this potential tussle with civil society groups
and those they represent can be avoided.

\setcounter{enumTemp}{\theenumi}
\end{enumerate}

\subsection{Role of identity}

\begin{enumerate}
\setcounter{enumi}{\theenumTemp}

\item\textit{Page 11: ``digital pounds are recorded anonymously on the Bank's
core ledger''.}  If the owner of digital assets is able to find them on the
ledger, then those assets must be linked to the identity of the owner.  Even if
they are recorded anonymously, the fact that the user can find them on the
ledger must reveal, either to the ``wallet'' service provider or to the ledger
operator, information linking the user to those assets.

\item\textit{Page 12: ``using their wallet to see their balance''.} Is the
balance recorded as a balance, rather than tokens?  If there is a balance on
the ledger that is updated during a transaction and accessed again in the
following transaction, then those two transactions are linked.  If two or more
transactions are linked, then information that can be used to identify the
owner is revealed.

\item\textit{Page 32: ``improved functionality for users, such as
programmability''.}  The examples given in the Consultation Paper of
user-specified rules to limit their own spending on certain products or to
facilitate saving are certainly achievable with a CBDC.  However, there are
important limits to programmability.  Rules that serve to bind a user to an
agreement with a third party, as well as rules that serve to restrict a user
from conducting certain kinds of transactions, must necessarily index the
identity of the user, implicitly or explicitly, for example via an account-like
custodial relationship or by embedding identity information, and thus
accountability, into individual tokens.  Such use cases undermine both consumer
privacy and the fungibility of the assets in question, and expose consumers to
human rights abuses akin to those that characterised the Australian cashless
welfare card~\cite{dss-au2022}.

\item\label{i:atm-limits}\textit{Page 40: ``In periods of banking or financial
stress... demand for digital pounds could be particularly strong''.}  During
such periods, the availability of digital pounds can be managed on a per-user
basis, just as the availability of cash can be managed, by limiting the rate of
receipt or withdrawals~\cite{ap2018}.  With a suitable system design, it is
likely that digital pounds can be managed even more easily.

\item\label{i:deposits}\textit{Page 41: ``increased risks... from banks
becoming more reliant on wholesale funding and less on deposits''.}  It is
possible to limit withdrawals per unit time on a per-user basis, in a manner
that mitigates the risk that individual users would convert large volumes of
bank deposits to CBDC.  See Item~\ref{i:receipt}.  However, a banking system
that implicitly requires consumers to not hold cash directly, because it
depends upon consumers to hold all of their current assets in custodial
accounts, is not particularly robust or sustainable.

\item\textit{Page 44: ``more susceptible to more frequent inflows and outflows
because it could be easier to switch into digital pounds than into cash''.} The
argument that the digital nature of CBDC implies that CBDC is harder to control
is false.  The rate of outflows can be managed by imposing user-based
restrictions on receipt and withdrawal of CBDC, which can be combined with
expiration policies and managed more efficiently than cash.  See
Item~\ref{i:receipt}.

\item\textit{Page 70: ``these data would be anonymised and not be considered
personal data''.}  It is not possible to anonymise personal data associated
with transactions without unlinking the transactions and assets from each
other.  Depending upon how the core ledger is implemented, including the
question of whether assets are represented as tokens or balances, unlinking the
transactions and assets from each other might not be possible.

\item\textit{Page 71: ``law enforcement agencies and competent authorities
could only access digital pound data where there is a fair and lawful basis''.}
This describes an ``exceptional access mechanism''; see
Item~\ref{i:exceptional}.  Presumably, authorities would request data from
regulated service providers, although if the CBDC were a token rather than a
balance and the consumer were to have direct possession and control, then the
consumer could (using privacy-enhancing technology such as blind signatures or
ZKP) furnish it to a vendor without revealing his or her identity.  Although
the payer would be anonymous, the recipient would be transparent.

\item\label{i:identity-assurance}\textit{Page 72: ``tiered identity
verification''.}  The essence of tiered identity verification is substantively
similar to a design described by the People's Bank of China in 2021, wherein
different categories of accounts are associated with different degrees of
identity assurance~\cite{pboc2021}.  Unfortunately, a lower degree of identity
assurance does not imply a higher degree of privacy.  The PBOC proposal for
tiered accounts is tantamount to a proposal for pre-paid debit cards with
varying information requirements~\cite{goodell2021}.  Unfortunately, consumers
stil lack privacy, since their transactions are linked through the repeated use
of the same account.  As a result, such an architecture not only fails to
deliver privacy but also encumbers regulatory compliance~\cite{goodell2022}.

\item\textit{Page 73: ``providers must collect the data that is required for
legal purposes''.}  Irrespective of the level of identity assurance for the
data used to establish an account, the data associated with multiple uses of an
account will reveal information that can be used to expose consumers to the
risk of profiling on the basis of the of the account.  See
Item~\ref{i:identity-assurance}.

\item\label{i:sponsor}\textit{Page 76: ``Non-resident access''.}  In a system
wherein one party must be transparent in every transaction, it is possible to
allow a sponsor to associate his or her identity with the CBDC asset in
exchange for furnishing the asset to an unidentified second party.  The second
party can then conduct transactions without revealing his or her identity, as
the sponsor normally would have done, except that the identity of the sponsor
would be associated with the payment transaction as the payer.  This can be
used to ensure accountability while allowing parties without identification
credentials suitable for AML/KYC procedures to receive and use CBDC.

\item\textit{Page 76: ``ensure that UK standards of resilience, consumer
protection, AML, KYC and any other legal requirements are upheld''.}  In most
circumstances, it makes sense that consumers would be required to satisfy
AML/KYC requirements to receive or withdraw CBDC in the first instance, thus
ensuring that CBDC is not commonly received or used by parties subject to
sanctions or conditions.  However in-person money changing services and cash
points with human operators could be permitted to waive the requirement to
provide documentation for small-value exchanges.

\setcounter{enumTemp}{\theenumi}
\end{enumerate}

\subsection{Role of the ledger}

\begin{enumerate}
\setcounter{enumi}{\theenumTemp}

\item\label{i:pass-through}\textit{Page 11: ``Private sector companies [would
offer] digital `pass-through' wallets to end users''.}  Functionally, a
``pass-through'' wallet seems to be a tool for accessing CBDC ``on the Bank's
core ledger'', rather than a device for storing digital assets.  This contrasts
with the definition of the term \textit{wallet} established by
experts~\cite{iso22739}, and it implies that users do not really have
possession of their money.  The use of a core ledger to store digital assets
also implies that banks do not have possession of their money, either.
Possession is centralised, even whilst the mechanism for accessing the tokens
is intermediated.

\item\textit{Page 13: ``the digital pound would... not make use of the same
energy-intensive technologies that underpin some cryptoassets''.}  Some
cryptoassets certainly make use of energy-intensive technologies.  For example,
cryptocurrencies that use permissionless ledgers often rely upon
computationally expensive mechanisms, such as proof of work, to defend against
Sybil attacks, and cryptocurrencies that make use of multiparty computation,
such as those implemented on the Ethereum platform, often entail
computationally expensive smart contracts.  However, the fact that some
cryptoassets that use distributed ledgers are environmentally harmful does not
imply that the use of a distributed ledger would necessarily be environmentally
harmful.  In particular, permissioned distributed ledgers that avoid multiparty
computation have a much smaller energy footprint, and models that avoid
recording every asset or transaction directly on the ledger are even more
energy-efficient.

\item\textit{Page 19: ``Blockchain technology... introduced digital assets
supported and distributed in a peer-to-peer fashion, backed by cryptography
alone''.}  The purpose of Blockchain technology is to establish an immutable
ledger via consensus among independent actors.  Blockchain technology can also
support the validation of transactions involving assets that are not
distributed (or redeemed) in a peer-to-peer fashion.  In addition, it is more
accurate to say that cryptocurrencies are \textit{enabled}, not
\textit{backed}, by cryptography; the ``backing'' is implicit to the
willingness of marketplaces (or specific traders) to accept cryptocurrencies as
a means of exchange.

\item\textit{Page 23: ``unbacked cryptoassets... are not considered an
efficient medium of exchange''.}  Inefficiency as a medium of exchange is not
intrinsic to unbacked cryptoassets; cryptocurrencies with this characteristic
typically either (a) use permissionless ledgers, which require computationally
expensive security measures to protect the integrity of the ledger; (b) manage
tokens directly on the ledger, thereby introducing substantial computational,
network, and storage overhead; (c) rely upon multiparty computation, for
example to execute smart contracts; or (d) some combination thereof.  Unbacked
cryptoassets of the future are likely to be more efficient, particularly with
oblivious transfers and self-validating assets, which dramatically reduce the
requisite communication and storage requirements; privacy-enhancing
technologies such as blind signatures and anonymising proxies can make it
easier to run a permissioned network without being shut down; and scalable
architectures that separate transaction processing from issuance, allowing a
broad base of issuers that can avoid direct involvement in
transactions~\cite{goodell2022}.

\item\textit{Page 24: ``a public digital pound infrastructure available to all
eligible private-sector firms that wish to develop new payment services''.}
There are several questions here.  First, what is the process for expressing a
wish to develop new payment services?  Second, what will delineate the
infrastructure from consumer-facing services?  We could imagine that
infrastructure might refer to distribution and redemption channels through
banks and other money service businesses, although we could also imagine that
the infrastructure refers to the operation of the ledger itself.  In this
system design, must retail users interact directly with the regulated operators
of a ``core ledger'', or can they also satisfy the requirements to execute
transactions by interacting with the ledger only indirectly, through an
unregulated service provider that interacts with a regulated operator of the
``core ledger''?  The distinction is important, because accountability for
system operators is non-negotiable.  See Item~\ref{i:rules}.

\item\textit{Page 53: ``a payment made in digital pounds between two users
would be processed and settled by a transfer on the Bank's core ledger''.}  The
assumption that assets should be directly recorded on the ledger is not
justified.  The assumption that the ledger should be controlled by a central
operator, which can exercise the option to equivocate or change the rules
without warning, is not justified.  See Item~\ref{i:rules}.

\item\textit{Page 53: ``The private sector... would provide digital
pass-through wallets''.}  For users to have possession and control of their own
money, wallets must not be ``provided''.  See Item~\ref{i:pass-through}.

\item\textit{Page 53: ``the core ledger operated by the Bank... might use
distributed ledger technology''.}  The core ledger should indeed use
distributed ledger technology, but mainly to enforce accountability of system
operators, to prevent equivocation, and to prevent a central operator from
changing the rules or history without warning.  See Item~\ref{i:rules}.

\item\textit{Page 56: ``pass-through wallets would allow users to hold and use
the digital pound''.}  This is not true with the architecture proposed in the
Consultation Paper, wherein the assets are held on the ledger and accessed only
via PIPs.  See Item~\ref{i:pass-through}.

\item\label{i:account}\textit{Page 56: access to digital pounds, make payments,
view balances and transaction history, mobility.}  The functions of ``wallets''
listed in the Consultation Paper are really the functions of \textit{accounts},
and the use of the term ``wallet'' is misleading.  The Consultation Paper seems
to envision a ``wallet'' as a kind of account.  See Items~\ref{i:provided}
and~\ref{i:directly}.

\item\textit{Page 59: ``Activities of Payment Interface Providers might need to
be restricted to safeguard system resilience''.}  Presumably, PIPs operating a
DLT system trusted by the central bank would be subject to regulations and
restrictions, just as regulators oversee the operation of decentralised
best-execution networks.  However, this does not imply that every PIP must
participate in the DLT system.  For example, with architectures that use
self-validating tokens, some consumer-facing PIPs could send updates to
regulated PIPs without handling private data or participating in ledger
consensus.

\setcounter{enumTemp}{\theenumi}
\end{enumerate}

\subsection{Role of the issuer}

\begin{enumerate}
\setcounter{enumi}{\theenumTemp}

\item\label{i:central}\textit{Page 11: ``the wallet simply passes instructions
from the user to the [Bank's] core ledger''.}  In this design, the Bank of
England operates a platform.  No justification is provided for requiring the
Bank to \textit{operate} the ledger rather than to \textit{oversee} a ledger
operated by private-sector businesses.  In addition to more closely resembling
the manner in which electronic payments are processed today, a decentralised
network would provide critical advantages in terms of security, availability,
and trust.

\item\label{i:rules}\textit{Page 11: ``The Bank would provide... central
infrastructure''.} Because the core ledger is not decentralised, its operator
can change the rules without asking, or substitute one version of history for
another.  This design characteristic also implies a tremendous incentive for
attackers to seize this point of control.

\item\label{i:stablecoins}\textit{Page 21: ``stablecoins used as a money-like
instrument should have... the ability to redeem at par in fiat''.}  It is
possible to extend protections similar to deposit insurance to
privacy-respecting, identity-unlinked stablecoins using a self-validating token
architecture~\cite{goodell2022}.  In the event of default by the issuer,
consumers holding stablecoins would be able to link their unblinded tokens to
the blinded token associated with their withdrawal transaction.  Recipients
holding stablecoins have their identities embedded into the tokens.  Recipients
without bank accounts with the issuer should opt to receive its stablecoins
directly into its bank account, which would be able to immediately redeem them
in exchange for a regulated bank transfer from the issuer.

\item\textit{Page 28: ``the digital pound would reduce the incentive to use
such non-sterling money''.}  The incentive to use non-sterling money might be
strong if users cannot possess or control sterling money directly.

\item\textit{Page 36: ``the digital pound would be exposed to risks of
electricity outages and cyber-attacks''.}  Risks to availability are amplified
by a design that requires direct involvement of the issuer or its services in
the payment channel.  See Item~\ref{i:central}.

\item\textit{Page 45: ``systemic stablecoin issuance would need to be fully
backed with high-quality and liquid assets''.}  See Item~\ref{i:stablecoins}.

\item\textit{Page 53: ``the digital pound should be designed as a platform
model, as originally set out in the Bank's 2020 Discussion Paper''.}  The
architecture described in the most recent (2023) Consultation Paper introduces
few fundamental changes relative to the architecture proposed in earlier
Consultation Papers, despite preponderant criticism of the platform model and the
proposition that the Bank should build and operate a ``core ledger''.  See
Item~\ref{i:central}.

\item\textit{Page 54: ``Given the possible single point of failure risk with
the platform model, it would be necessary to ensure the infrastructure is
protected to the very highest standards''.}  The single point of failure in the
proposed model indeed represents a major weakness that can be avoided.  The
core ledger is a high-value target that will be expensive, perhaps impossible,
to adequately protect, and critical infrastructure can easily be compromised
without detection.  A far better approach would be to oversee a decentralised
network without operating it directly.  The model wherein a decentralised
network is overseen by a regulator is already used in financial infrastructure
in the context of best-execution networks~\cite{nms-changes} and works well.
See Item~\ref{i:central}.  It is dangerous for a single party to be in control
of the \textit{de facto} history of all transactions.  This has never been the
case before and is unjustified.

\item\textit{Page 55: ``The core ledger would be a single piece of
infrastructure''.}  It would be better from a security, availability, and
accountability perspective for the core ledger to be a network of service
providers and should not record individual assets directly, either in the form
of UTXO systems, such as Bitcoin, or multiparty computation systems, such as
Ethereum.  See Item~\ref{i:central}.

\setcounter{enumTemp}{\theenumi}
\end{enumerate}

\section{Responses to consultation questions}
\label{s:consultation}

\begin{enumerate}

\item\textit{Do you have comments on how trends in payments may evolve and the
opportunities and risks that they may entail?}

The following scenarios are possible and somewhat likely to emerge as
the payments landscape continues to devolve over the next decade:

\begin{enumerate}

\item increasing exclusion of cash-paying consumers in many circumstances;

\item an economic and regulatory context that favours and incentivises data
collection and surveillance by custodians;

\item continued support for custodial models for consumer money by governments
in preference to privately-held money, particularly in response to lobbying
efforts by incumbent stakeholders;

\item support from governments for public payments infrastructure that enables
data harvesting, using the implicit data subsidy as a means to reduce costs;

\item over-compliance with AML/CTF rules, leading to unnecessary surveillance
and deepening links between consumer purchases and personal data belonging to
consumers, thus increasing the commercial value of the data harvesting
opportunity;

\item an increasing use of money laundering techniques that do not rely upon
the use of regulated payment systems at all, coupled with continued high cost
and low benefit to the use of suspicious activity reports;

\item the emergence of alternative payment solutions operated by the private
sector, both domestically and internationally, for use in the UK;

\item the emergence and proliferation of ``outside solutions'' allowing
consumers to make digital payments using assets that they possess and control,
outside custodial relationships, with revenues for operating the infrastructure
used by such solutions accruing to criminal organisations in case they are not
supported by national governments;

\item self-certifying assets, including non-fungible tokens, that are not
stored or managed directly on a ledger but instead make use of the ledger for
verifying integrity and validity; and

\item oblivious transfers, wherein service providers that facilitate
transactions know nothing about the contents or parties to the transactions
themselves.

\end{enumerate}

\item\textit{Do you have comments on our proposition for the roles and
responsibilities of private sector digital wallets as set out in the platform
model? Do you agree that private sector digital wallet providers should not
hold end users' funds directly on their balance sheets?}

The concepts ``wallet'', ``account'', and ``address'' are often confused.  A
wallet is not something that is or can be ``provided'', or that wallets are
containers that are held or controlled by a third party on behalf of their
users.  Something fitting that description might be more accurately described
as an ``account''.

It is a dangerous mistake to think of a wallet in such terms.  A wallet is
correctly defined not as a service, but as an ``\textit{application or
mechanism} used to generate, manage, store or use'' digital
assets~\cite{iso22739} (emphasis added).  They are intended to operate at the
behest of their users, and there is no reason for users to be limited to the
type or number of wallets that they use.

Wallets do not operate at the behest of their users if they are created by
third parties, or registered with third parties, or are recognisable across
multiple transactions, all of which imply that third parties can link the
various actions taken by users of the same wallet over time, which implies that
the wallet serves those parties, not their users.

So, for users to have privacy, as well as possession and control of their
assets, their wallets:

\begin{enumerate}

\item must not contain private keys issued by an authority, since such keys
would be under the control of the authority;

\item must not be registered with an authority, since wallet identification
across multiple transactions allows the user to be profiled;

\item must not be identifiable across multiple transactions, since a profile can
be built by counterparties or other observers even without the involvement of
an authority;

\item must not be required to contain 'trusted computing' modules or 'secure
enclaves' that can perform remote attestations, thus constraining what the user
can do with assets in the wallet; and

\item should not normally be held by custodians, lest the custodians engage in
profiling, discrimination, or gatekeeping against the interests of their users.

\end{enumerate}

Wallet identification in the general case is dangerous, since it can be used to
restrict how users use their assets.  A true digital wallet, like a true
physical wallet, is completely under the control of its user and reveals
nothing during transactions of the assets it contains other than the assets
themselves.

\item\textit{Do you agree that the Bank should not have access to users'
personal data, but instead see anonymised transaction data and aggregated
system-wide data for the running of the core ledger? What views do you have on
a privacy-enhancing digital pound?}

The bank should not have access to users' personal data, although the personal
data that are most important to avoid collecting concern the payer associated
with a particular transaction, or the set of transactions associated with some
particular payer.  Personal data may be required in the process of receiving or
withdrawing CBDC to satisfy AML/KYC requirements, and such data may be
collected by money services businesses and shared with regulators and
authorities.  However, the main risk to the privacy of individual persons
results from the use of transaction information to create profiles of
consumers, for example as ``alternative data'' in the analysis or
categorisation of an individual on the basis of individual habits, location,
decisions, and modes of life.  Since there is no way for consumers to verify
that data are not used once collected, consumer privacy critically depends upon
not linking consumers to transactions that reveal how they spend their money.
Specifically, individual payers must have some form of technical guarantee that
their transactions can never be linked to them, by any party, including not
only the Bank, but also system operators and authorities, without both their
specific authorisation and unforced consent.·

Regulators and authorities should have technical access to all of the
information to which they legitimately have access, which does not generally
include the identity of payers in transactions.  There are valid reasons,
including tax and anti-fraud regulations, for which at least one party to every
transaction should be known to regulators and authorities, and some proposed
CBDC architectures achieve this without exposing the identity of the consumer
in the general case~\cite{bis2022,goodell2022}.  We envision that most
consumers would be known to regulators and authorities when they receive or
withdraw CBDC, but their identities would not be linkable to their spending
transactions.  We also envision that regulators and authorities would have
access to the identity of the recipient of most CBDC payments.

The Bank should certainly have access to aggregated system-wide data, as well
as anonymised transaction data.  (Authorities and regulators would generally
have access to the identity of the recipient in most transactions, but the Bank
would not require access to that information.)  It is less obvious, however,
that this information should be stored on a core ledger; storing the data with
the CBDC assets themselves would make the operation of the payment system more
scalable~\cite{goodell2022}.

CBDC in the UK must be privacy-enhancing, and we expect that this means that it
would make use of privacy-enhancing technologies such as Chaumian blind
signatures~\cite{chaum1982} or zero-knowledge proofs to prevent retail
consumers from being linked to their payments against their wishes.  ``Privacy
by promise'' is no substitute for privacy by design.

\item\textit{What are your views on the provision and utility of tiered access
to the digital pound that is linked to user identity information?}

The proposal of ``tiered access'' is substantively similar to a proposal by the
People's Bank of China in its 2021 discussion paper~\cite{pboc2021}.  For our
view on the proposal in the 2023 Consultation Paper by the Bank of England and HM
Treasury, we reference a contemporaneous analysis of the 2021 PBOC paper:

\begin{quote}

``Consumers have a right to conduct low-risk transactions with merchants,
vendors, and other providers of retail goods and services, without revealing
personal information that can be used to associate themselves with the
transaction.''~\cite{goodell2021}

``To the extent that successive transactions with the same account constitute
linked attributes about a user, the user is pseudonymous and not anonymous...
We believe that the salient question is how much privacy we want ordinary
persons to have, and with `managed anonymity', the answer is not much.
Inasmuch as transactions use accounts, which might be described as `wallets'
(in contrast to non-custodial wallets), the successive transactions that
pseudonymous users make using such accounts would be linked to each other,
hence the author's reference to a `loosely-coupled account linkage'.  So, while
a consumer could in principle use an anonymous account to make a single
payment, liquidating the entire account, most typical use cases would see
consumers acting as price-takers, so an exact match between the price of an
item and the value in an account would be unlikely.  The user of an account,
having completed a purchase of value less than the value of the account, would
then either use the account again for a subsequent transaction or forfeit the
difference between the agreed price and the value of the account.  As a result,
we might conclude that while malicious actors could use anonymous accounts to
achieve a reasonable level of anonymity, anonymous accounts would afford much
less privacy to ordinary, non-nefarious users.''~\cite{goodell2021}

``The authors of the PBOC report lightly touch on this final issue through
their discussion of `managed anonymity' whereby `the PBOC sets up a firewall
for e-CNY-related information, and strictly implements information security and
privacy protocols, such as designating special personnel to manage information,
separating e-CNY from other businesses, applying a tiered authorization system,
putting in place checks and balances, and conducting internal
audits'~\cite{pboc2021}.  Presumably, the objective of such processes would be
to discourage `arbitrary' access to that information by `authorized
operators', which it identifies as primarily comprising `commercial banks and
licensed non-bank payment institutions'~\cite{pboc2021} that circulate the
e-CNY to the public.  However, as the entities that the authors call
`wallets' are essentially accounts, it is not obvious that the data
protection mechanisms recommended by the authors can achieve their stated
goals, and, given the system-level implications highlighted, the role of system
operators as arbiters of such mechanisms underscores the trade-off that is
inherently being made among the guiding principles.''~\cite{goodell2021}

\end{quote}

A better approach to privacy would focus on the discoverability of links among
transactions or between transactions and user data, since those links form the
basis for identification and the potential for profiling a user on the basis of
payments.  Low-value, relatively safe transactions should be fully unlinked,
wherein no identity information is provided for the payer.  For high-value
transactions, or transactions that are considered sensitive because of their
nature, the recipient may be required to demand identification as a condition
for redemption of the CBDC asset.  It may be expected that consumers would be
required to provide identification when receiving direct CBDC payments or when
withdrawing CBDC assets from bank accounts, which would presumably also require
routine AML/KYC procedures.  It may be expected that consumers would not be
required to provide identification when exchanging limited amounts of cash for
CBDC, and perhaps also when exchanging the CBDC issued by one central bank for
the CBDC issued by another central bank.

\item\textit{What views do you have on the embedding of privacy-enhancing
techniques to give users more control of the level of privacy that they can
ascribe to their personal transactions data?}

Techniques that afford users more control over the collection of their data are
certainly useful.  It is indeed promising that the proposal specifies the use
of privacy by design rather than requiring users to accept data collection,
after which claims about the storage and use of data cannot be verified.

However, we note that the primary concern for users is less about extent of the
personal data required to satisfy AML/KYC requirements, and more about the use
of transaction information that associates consumers with how they spend their
money.  It is not clear that the proposal described in the Consultation Paper
offers sufficient protection (i.e., privacy by design) to prevent the ability
to link a transaction to an individual payer or to link a set of multiple
transactions to the same payer.  We argue that it is necessary to first affirm
the right of individual persons to freely engage in the digital economy, with
privacy by default and without requiring the consumer to be linked to the
transaction in most small-value payments.  We imagine that privacy-enhancing
technologies, such as blind signatures and zero-knowledge proofs, will be
necessary to achieve this property, along with common-sense mechanisms, such as
leveraging the network connectivity of the recipient and not requiring wallets
to be identified, to reduce the risk of metadata leakage.

\item\textit{Do you have comments on our proposal that in-store, online and
person-to-person payments should be highest priority payments in scope? Are any
other payments in scope which need further work?}

Yes, in-store, online, and person-to-person payments should have the highest
priority, since these payments generally form the basis of consumer
transactions.  Nevertheless, there are some other payment types that should be
considered important as well, including but not limited to:

\begin{enumerate}

\item\textit{Payments from businesses to individuals.}  Notwithstanding the
fact that most consumers would probably prefer to receive payments into a bank
account rather than a wallet (Item~\ref{i:salary}), it should be possible for
individuals to receive salaries, wages, or tips from businesses into their
non-custodial wallets.  Under such circumstances, a business would most likely
be expected to embed information identifying itself into the transaction, and
the consumer might be expected to deposit the CBDC into a bank account before
spending it.

\item\textit{Transfers to sponsored recipients.}  In situations wherein a
consumer is unable to provide sufficient documentation to satisfy AML/KYC
criteria necessary to receive CBDC, it should be possible for a sponsor to
transfer CBDC to this consumer by embedding information identifying the sponsor
into the transaction.  See Item~\ref{i:sponsor}.  Note that this case differs
from the case of payments from businesses to individuals in that sponsored
consumers would be expected to spend the CBDC that they receive via a
sponsoring transaction.

\item\textit{Exchanges of cash or foreign CBDC for domestic CBDC.}  Consumers
should be able to exchange small quantities of cash or foreign CBDC for
domestic CBDC in-person, without providing documentation or identity
information.

\item\textit{Exchanges of domestic CBDC for cash or foreign CBDC.}  Consumers
should be able to exchange small quantities of domestic CBDC for cash or
foreign CBDC, without providing documentation or identity information.
(Additional requirements related to receiving foreign CBDC might be specified
by the foreign issuer.)

\end{enumerate}

\item\textit{What do you consider to be the appropriate level of limits on
individual's holdings in transition? Do you agree with our proposed limits
within the £10,000--£20,000 range?  Do you have views on the benefits and risks
of a lower limit, such as £5,000?}

There should not be a limit on the aggregate holdings of individual consumers.
Note that there is no limit on how much cash a consumer can own, and to protect
this right for the digital age, we must ensure that there is no
\textit{explicit} limit on the aggregate volume of digital cash that a consumer
can own.  Instead, there should be a limit on \textit{withdrawals}, implemented
by requiring banks and money services businesses to enforce the limit.  A limit
on withdrawals can be used in conjunction with expiration rules to define an
\textit{implicit} limit on the aggregate volume of CBDC that a consumer can
own, although the limit on withdrawals is likely to be more flexible and more
effective in mitigating the risk of financial instability (see
Item~\ref{i:instability}).  Limits can be imposed as a maximum quantity that a
consumer can receive or withdraw over some period of time, or as a maximum
quantity that a consumer can withdraw without correspondingly making deposits,
or both.  Limits can also be proposed on a per-user basis; some consumers might
have more restrictive limits than others.  Finally, limits can be dynamically
adjusted in response to market conditions or circumstances.

In all cases, to avoid curtailing the right of consumers to privacy, we must
ensure that the limit for the quantity of CBDC that a consumer can spend
without providing identifying information must be not less than the quantity of
cash that a consumer can spend without providing identifying information, and
this quantity must increase with inflation to prevent a \textit{de facto}
curtailment of the right to privacy over time.  Since private payments
implicitly require users to hold money in non-custodial CBDC wallets,
non-custodial CBDC wallets must be capable of receiving and holding at least
this quantity.  Ultimately, however, authorities should not impose restrictions
on the wallets themselves, even as they may impose certain restrictions on
receipt and withdrawal of funds into wallets.

\item\textit{Considering our proposal for limits on individual holdings, what
views do you have on how corporates' use of digital pounds should be managed in
transition? Should all corporates be able to hold digital pounds, or should
some corporates be restricted?}

There are several salient aspects to the question of how the use of CBDC by
corporations should be limited:

\begin{enumerate}

\item whether there should be limits on how much CBDC corporations can receive
in the form of payments;

\item whether there should be limits on how much CBDC corporations can withdraw;

\item how limits should be implemented; and

\item whether the same limits should apply to all corporations.

\end{enumerate}

We imagine that limitations on the receipt and withdrawal of CBDC can apply to
corporations, just as it can apply to individual persons.  The limitations can
take the form of restrictions on how much CBDC a corporation can receive or
withdraw over a period of time, restrictions on the quantity of CBDC that a
corporation can receive or withdraw net of the quantity of CBDC that a
corporation has deposited, or both.  We imagine that different limitations, as
well as different restrictions on conditions for accepting deposits of CBDC,
might apply to different corporations.  We also imagine that some corporations
might be allowed to receive CBDC directly into non-custodial wallets, whereas
others might be rqeuired to receive CBDC into custodial accounts managed by
regulated intermediaries such as banks.

\item\textit{Do you have comments on our proposal that non-UK residents should
have access to the digital pound, on the same basis as UK residents?}

Non-UK residents should have access to the digital pound for several reasons,
including but not limited to the following:

\begin{enumerate}

\item\textit{Non-UK residents visit the UK from time to time.}  Visitors will
need a public payment option to engage with the digital economy, just as
residents do, particularly for use cases in which cash payments are
discouraged, refused, or considered impractical.

\item\textit{Non-UK residents seek to purchase goods and services from UK
vendors.} Under normal circumstances, non-UK residents require a public payment
option to purchase goods and services from UK vendors via the Internet or over
the phone, whether or not they are visiting the UK.  UK-issued CBDC can
facilitate such transactions.

\item\textit{UK vendors seek to market goods and services to non-UK residents.}
UK vendors might require or prefer a public payment option for receiving
payments, including from non-UK residents.

\end{enumerate}

So, we conclude that non-UK residents should have a means to receive CBDC into
non-custodial wallets, just as UK residents should have a means to do so.
Nevertheless, it is not entirely obvious that UK residents and non-UK residents
would share the same user journey for UK-issued CBDC.  For example, non-UK
residents might have a different set of requirements for furnishing AML/KYC
documentation, or in certain circumstances they might require sponsorship (see
Item~\ref{i:sponsor}).  If their use of CBDC mainly concerns purchasing goods
and services from UK e-commerce vendors, then a means of acquiring UK-issued
CBDC might be facilitated by other e-commerce service providers.

\item\textit{Given our primary motivations, does our proposed design for the
digital pound meet its objectives?}

Unfortunately, no.  Sections~\ref{s:summary} and~\ref{s:detailed} explain why.

\item\textit{Which design choices should we consider in order to support
financial inclusion?}

Financial inclusion depends upon many factors.  Design choices supporting
financial inclusion include:

\begin{enumerate}

\item\textit{True privacy by design}, wherein consumer transactions are not
linkable to the identity of the payer or to each other in the general case,
most likely requiring privacy-enhancing technologies such as blind
signatures;

\item\textit{Non-custodial wallets}, offering true possession and control to
owners of CBDC assets, without technical limits on aggregate holdings, and
which can be held directly by the owners without the involvement of third
parties;

\item\textit{The use of tokens rather than balances}, to avoid implicit
identity as an emergent property of the use of balances;

\item\textit{Self-validating assets}, which can be held off-ledger and use the
ledger only to ensure their validity and integrity, to minimise the role of
gatekeepers; and

\item\textit{A decentralised ledger}, operated by a set of independent
private-sector businesses overseen by the central bank, rather than a
centralised ledger directly operated by the central bank, to ensure that access
is not limited by the technical or practical limitations of the central bank.

\end{enumerate}

Sections~\ref{s:summary} and~\ref{s:detailed} offer a detailed explanation in
support of these design choices.

\item\textit{The Bank and HM Treasury will have due regard to the public sector
equality duty, including considering the impact of proposals for the design of
the digital pound on those who share protected characteristics, as provided by
the Equality Act 2010.  Please indicate if you believe any of the proposals in
this Consultation Paper are likely to impact persons who share such protected
characteristics and, if so, please explain which groups of persons, what the
impact on such groups might be and if you have any views on how impact could be
mitigated.}

The proposals described in the Consultation Paper are likely to impact persons
with protected characteristics for reasons related to privacy and
accessibility.  Specifically, because the proposed system is not private by
design, the following groups of individuals could be exposed to the risk of
profiling, discrimination, blackmail, or violence:

\begin{enumerate}

\item victims of abuse;

\item witnesses to crime;

\item refugees;

\item stateless persons;

\item persons who are members of protected groups;

\item persons who are exposed to discrimination on the basis of protected
traits, such as race, gender, religion, disability, age, sexual orientation,
veteran status, and so on;

\item persons with pre-existing medical conditions;

\item persons who have been blacklisted as a result of a history of criminal
convictions or financial default; and

\item persons in abusive personal, domestic, or business relationships.

\end{enumerate}

In addition, because the proposed system relies upon service provision by third
parties and does not offer an open architecture for non-custodial wallets,
there is a risk that third-party business models might not provide adequate
support for certain groups, such as:

\begin{enumerate}

\item persons with vision impairment;

\item persons who cannot afford smartphones;

\item undocumented persons;

\item homeless persons; and

\item persons whose possessions are under the control of another person.

\end{enumerate}

\end{enumerate}

\section*{Acknowledgements}

We thank Professor Tomaso Aste for his ongoing support of projects in this
area.  Geoff Goodell is also an associate of the Systemic Risk Centre of the
London School of Economics as well as the UCL Centre for Blockchain
Technologies.  We also acknowledge EPSRC and the PETRAS Research Centre
EP/S035362/1 for the FIRE Project.


\begin{thebibliography}{1}\raggedright
\footnotesize
\bibitem{boe2023}{
    Bank of England and HM Treasury.
    ``The digital pound: a new form of money for households and businesses?''
    Discussion paper,
    February 2023.
    [online]
    \url{https://www.bankofengland.co.uk/-/media/boe/files/paper/2023/the-digital-pound-consultation-working-paper.pdf}
    [retrieved 2023-04-01]
}
\bibitem{boe2020}{
    Bank of England.
    ``Central Bank Digital Currency: opportunities, challenges and design.''
    Discussion paper,
    2020-03-12.
    [online]
    \url{https://www.bankofengland.co.uk/paper/2020/central-bank-digital-currency-opportunities-challenges-and-design-discussion-paper}
    [retrieved 2020-03-16]
}
\bibitem{boe2021}{
    Bank of England.
    ``New forms of digital money.''
    Discussion paper,
    2021-06-07.
    [online]
    \url{https://www.bankofengland.co.uk/paper/2021/new-forms-of-digital-money}
    [retrieved 2021-07-05]
}
\bibitem{allix2019}{
    J Allix and F Aliyev.
    ``Cash versus Cashless: Consumers need a right to use cash.''
    Bureau Europ\'een des Unions de Consommateurs (BEUC),
    2019-09-25.
    [online]
    \url{https://www.beuc.eu/sites/default/files/publications/beuc-x-2019-052_cash_versus_cashless.pdf}
    [retrieved 2023-05-01]
}
\bibitem{buttarelli2019}{
    G Buttarelli.
    ``Privacy in a Cashless World.''
    European Data Protection Supervisor,
    2019-01-28.
    [online]
    \url{https://edps.europa.eu/press-publications/press-news/blog/privacy-cashless-world_en}
    [retrieved 2023-05-01]
}
\bibitem{ca-opc2016}{
    Office of the Privacy Commissioner of Canada.
    ``Electronic and digital payments and privacy.''
    2016-09-01.
    [online]
    \url{https://www.priv.gc.ca/en/privacy-topics/technology/mobile-and-digital-devices/02_05_d_68_dp/}
    [retrieved 2023-05-01]
}
\bibitem{edpb2021}{
    European Data Protection Board.
    ``EDPB letter to the European institutions on the privacy and data protection aspects of a possible digital euro.''
    2021-06-18.
    [online]
    \url{https://edpb.europa.eu/system/files/2021-07/edpb_letter_out_2021_0112-digitaleuro-toep_en.pdf}
    [retrieved 2023-05-01]
}
\bibitem{cnil2022}{
    Commission Nationale de l'Informatique et des Libert\'es (CNIL).
    ``Digital euro: what is at stake for privacy and personal data protection?''
    2022-02-14.
    [online]
    \url{https://www.cnil.fr/en/digital-euro-what-stake-privacy-and-personal-data-protection}
    [retrieved 2023-05-01]
}
\bibitem{mai2019}{
    H Mai.
    ``Cash empowers the individual through data protection.''
    Deutsche Bank Research,
    2019-07-02.
    [online]
    \url{https://www.dbresearch.com/PROD/RPS_EN-PROD/PROD0000000000495958/Cash_empowers_the_individual_through_data_protecti.xhtml}
    [retrieved 2023-05-01]
}
\bibitem{higgins2021}{
    E Higgins.
    ``What Are the Privacy Implications of a Cashless Society?''
    Lawpath,
    2021-06-29.
    [online]
    \url{https://lawpath.com.au/blog/what-are-the-privacy-implications-of-a-cashless-society}
    [retrieved 2023-05-01]
}
\bibitem{naacp2019}{
    National Association for the Advancement of Colored People (NAACP).
    ``Cashless Retail Transactions Promotes Discrimination in our Communities.''
    2019.
    [online]
    \url{https://naacp.org/resources/cashless-retail-transactions-promotes-discrimination-our-communities}
    [retrieved 2023-05-01]
}
\bibitem{gladstein2021}{
    A Gladstein.
    ``Financial Freedom and Privacy in the Post-Cash World.''
    Cato Journal,
    Cato Institute.
    Spring/summer 2021.
    [online]
    \url{https://www.cato.org/cato-journal/spring/summer-2021/financial-freedom-privacy-post-cash-world}
    [retrieved 2023-05-01]
}
\bibitem{kahn2018}{
    C Kahn.
    ``Payment Systems and Privacy.''
    Federal Reserve Bank of St Louis
    \textit{Review} \textbf{100}(4),
    Fourth Quarter 2018,
    pp. 337--44.
    [online]
    \url{https://doi.org/10.20955/r.100.337-44}
    [retrieved 2023-05-01]
}
\bibitem{bindseil2023}{
    U Bindseil and D Schneeberger.
    ``Cash or cashless? How people pay.''
    European Central Bank,
    2023-02-06.
    [online]
    \url{https://web.archive.org/web/20230421122856/https://www.ecb.europa.eu/press/blog/date/2023/html/ecb.blog230206~1ea270a762.en.html}
    [retrieved 2023-05-01]
}
\bibitem{garratt2021}{
    R Garratt and M van Oordt.
    ``Privacy as a Public Good: A Case for Electronic Cash.''
    \textit{Journal of Political Economy} \textbf{129}(7),
    July 2021.
    [online]
    \url{https://doi.org/10.1086/714133}
    [retrieved 2023-05-01]
}
\bibitem{coustick-deal2015}{
    R Coustick-Deal.
    ``Responding to ``Nothing to Hide, Nothing to Fear.''
    Open Rights Group,
    2015-12-04.
    [online]
    \url{https://www.openrightsgroup.org/blog/responding-to-nothing-to-hide-nothing-to-fear/}
    [retrieved 2022-07-13]
}
\bibitem{abelson1997}{
    H. Abelson, R. Anderson, S. Bellovin, J. Benaloh, M. Blaze, W. Diffie, J. Gilmore, P. Neumann, R. Rivest, J. Schiller, and B. Schneier.
    ``The Risks of Key Recovery, Key Escrow, and Trusted Third-Party Encryption.''
    \texttt{doi:10.7916/D8GM8F2W},
    1997-05-27.
    [online]
    \url{https://academiccommons.columbia.edu/doi/10.7916/D8R2176H/download}
    [retrieved 2019-03-11]
}
\bibitem{abelson2015}{
    H. Abelson, R. Anderson, S. Bellovin, J. Benaloh, M. Blaze, W. Diffie, J. Gilmore, M. Green, S. Landau, P. Neumann, R. Rivest, J. Schiller, B. Schneier, M. Specter, and D. Weitzner.
    ``Keys under doormats: mandating insecurity by requiring government access to all data and communications.''
    \textit{Journal of Cybersecurity} \textbf{1}(1),
    pp. 69--79,
    \texttt{doi:10.1093/cybsec/tyv009},
    2015-11-17.
    [online]
    \url{https://academiccommons.columbia.edu/doi/10.7916/D82N5D59/download}
    [retrieved 2019-03-11]
}
\bibitem{benaloh2018}{
    J. Benaloh.
    ``What if Responsible Encryption Back-Doors Were Possible?''
    Lawfare Blog,
    2018-11-29.
    [online]
    \url{https://www.lawfareblog.com/what-if-responsible-encryption-back-doors-were-possible}
    [retrieved 2018-12-11]
}
\bibitem{nissenbaum2017}{
    H Nissenbaum.
    ``Deregulating Collection: Must Privacy Give Way to Use Regulation?''
    May 2017.
    \url{https://doi.org/10.2139/ssrn.3092282}
}
\bibitem{rychwalska2021}{
    A Rychwalska, G Goodell, and M Roszczynska-Kurasinska.
    ``Management of Big Data in the public sector: System-level risks and design principles.''
    Presented at Conference on Complex Systems (CCS), Singapore, September 2019.
    \textit{Surveillance \& Society}, Vol 19 No 1 (2021), pp. 22-36.
    \url{https://doi.org/10.24908/ss.v19i1.13986}
}
\bibitem{bis2022}{
    Bank for International Settlements.
    ``Project Tourbillon explores cyber resiliency, scalability and privacy in a prototype CBDC.''
    [online]
    \url{https://www.bis.org/about/bisih/topics/cbdc/tourbillon.htm}
    [retrieved 2023-04-01]
}
\bibitem{goodell2022}{
    G Goodell, D Toliver, and H Nakib.
    ``A Scalable Architecture for Electronic Payments.''
    Presented at WTSC 2022, Grenada, May 2022.
    To appear, \textit{Lecture Notes in Computer Science} 13412.
    [online]
    \url{http://doi.org/10.2139/ssrn.3951988}
}
\bibitem{iso22739}{
    International Organisation for Standardization (ISO).
    \textit{ISO/DIS 22739
Blockchain and distributed ledger technologies -- Vocabulary.}
    Clause 3.99, \textit{wallet}.
    [online]
    \url{https://www.iso.org/obp/ui/#iso:std:iso:22739:dis:ed-2:v1:en:term:3.99}
    [retrieved 2023-05-01]
}
\bibitem{goodell2022a}{
    G Goodell.
    ``Certified Hardware Requirements Undermine Digital Currency.''
    Available at SSRN,
    2022-09-29.
    [online]
    \url{https://dx.doi.org/10.2139/ssrn.4388717}
}
\bibitem{dss-au2022}{
    A Rishworth MP, B Shorten MP, and J Elliot MP,
    Ministers for the Department of Social Services.
    ``Cashless debit card program to end following passage of legislation.''
    Media release,
    2022-09-28.
    [online]
    \url{https://ministers.dss.gov.au/media-releases/9221}
    [retrieved 2023-05-01]
}
\bibitem{staschen2022}{
    S Staschen and A Plaitakis.
    ``Platform-Based Finance: Are Regulators Up to the Data Protection Task?''
    Consultative Group to Assist the Poor,
    2022-03-01.
    [online]
    \url{https://web.archive.org/web/20230323180634/https://www.cgap.org/blog/platform-based-finance-are-regulators-to-data-protection-task}
    [retrieved 2023-03-23]
}
\bibitem{woolf2022}{
    M Woolf.
    ``RCMP gave banks police info on Ottawa protesters with a list of accounts to freeze.''
    The Canadian Press,
    2022-03-08.
    [online]
    \url{https://www.cbc.ca/news/politics/rcmp-names-banks-freeze-1.6376955}
    [retrieved 2023-05-01]
}
\bibitem{fraser2022}{
    A Fraser.
    ``RCMP shared Freedom Convoy `blacklist' across the world.''
    Ottawa Sun,
    2022-09-22.
    [online]
    \url{https://ottawasun.com/news/national/rcmp-sent-freedom-convoy-blacklist-to-lobbyists}
    [retrieved 2023-05-01]
}
\bibitem{auer2020a}{
    R. Auer and R. B\"ohme.
    ``The technology of retail central bank digital currency.''
    \textit{BIS Quarterly Review},
    March 2020,
    pp. 85--100.
    [online]
    \url{https://www.bis.org/publ/qtrpdf/r_qt2003j.pdf}
    [retrieved 2020-05-19]
}
\bibitem{goodell2021}{
    G Goodell and H Nakib.
    ``The Development of Central Bank Digital Currency in China: An Analysis.''
    LSE Systemic Risk Centre Opinion Piece,
    November 2021.
    [online]
    \url{https://www.systemicrisk.ac.uk/sites/default/files/2021-12/2108.05946.pdf}
}
\bibitem{bis2012}{
    Bank for International Settlements.
    ``Payment, clearing and settlement systems in the CPSS countries.''
    Committee on Payment and Settlement Systems ``Red Book'',
    Volume 2,
    November 2012.
    [online]
    \url{https://www.bis.org/cpmi/publ/d105.pdf}
    [retrieved 2020-05-31]
}
\bibitem{bis2012a}{
    Bank for International Settlements.
    ``Payment, clearing and settlement systems in the United Kingdom.''
    Committee on Payment and Settlement Systems ``Red Book'',
    Volume 2,
    November 2012,
    pp. 445--446.
    [online]
    \url{https://www.bis.org/cpmi/publ/d105_uk.pdf}
    [retrieved 2020-04-16]
}
\bibitem{goodell2021a}{
    G Goodell, H Nakib, and P Tasca.
    ``A Digital Currency Architecture for Privacy and Owner-Custodianship.''
    \textit{Future Internet} 2021, 13(5),
    May 2021.
    [online]
    \url{https://doi.org/10.3390/fi13050130}
}
\bibitem{tucker2023}{
    E Tucker, T Copp, and M Balsamo.
    ``Guardsman arrested in leak of classified military documents.''
    AP News,
    2023-04-14.
    [online]
    \url{https://apnews.com/article/leaked-documents-pentagon-justice-department-russia-war-d3272b34702d564fe07a480598bcd174}
    [retrieved 2023-05-01]
}
\bibitem{nms-changes}{
    U.S. Securities and Exchange Commission.
    ``SEC Proposes Improvements to Governance of Market Data Plans.''
    Press Release,
    2020-01-08.
    [online]
    \url{https://www.sec.gov/news/press-release/2020-5}
    [retrieved 2020-07-16]
}
\bibitem{access2019}{
    Access to Cash Review (UK),
    Final Report,
    March 2019.
    \url{https://www.accesstocash.org.uk/media/1087/final-report-final-web.pdf}
}
\bibitem{tischer2020}{
    D Tisher, J Evans, K Cross, R Scott, and I Oxley.
    ``Where to Withdraw?  Mapping access to cash across the UK.''
    University of Bristol,
    November 2020.
    \url{http://www.bristol.ac.uk/media-library/sites/geography/pfrc/Where%20to%20withdraw%20-%20mapping%20access%20to%20cash%20across%20the%20UK.pdf}
}
\bibitem{schumpeter1942}{
    J. Schumpeter.
    \textit{Capitalism, Socialism, and Democracy}.
    New York: Harper and Brothers,
    1942,
    pp. 90-91.
}
\bibitem{tumin2022}{
    R Tumin.
    ``Cash or Card for a Cone? Van Leeuwen Must Take Both, N.Y.C. Says.''
    The New York Times,
    2022-10-20.
    [online]
    \url{https://www.nytimes.com/2022/10/20/nyregion/van-leeuwen-cashless-ban-settlement.html}
    [retrieved 2022-10-21]
}
\bibitem{hrw-parallel}{
    S St Vincent.
    ``Dark Side: Secret Origins of Evidence in US Criminal Cases.''
    Human Rights Watch,
    2018-01-09.
    [online]
    \url{https://www.hrw.org/report/2018/01/09/dark-side/secret-origins-evidence-us-criminal-cases}
    [retrieved 2021-10-25]
}
\bibitem{armer1968}{
    P Armer.
    ``Privacy Aspects of the Cashless and Checkless Society.''
    Testimony before the US Senate Subcommittee on Administrative Practice and Procedure.
    1968-02-06, as published by the RAND Corporation, April 1968.
    \url{https://www.rand.org/content/dam/rand/pubs/papers/2013/P3822.pdf}
}
\bibitem{armer1975}{
    P Armer.
    ``Computer Technology and Surveillance.''
    \textit{Computers and People} 24(9),
    pp. 8--11,
    September 1975.
    \url{https://archive.org/stream/bitsavers_computersA_3986915/197509#page/n7/mode/2up}
}
\bibitem{fairfield2015}{
    J Fairfield and C Engel.
    ``Privacy as a Public Good.''
    Duke Law Journal \textbf{65}(385),
    2015.
    [online]
    \url{http://scholarship.law.duke.edu/dlj/vol65/iss3/1}
    [retrieved 2021-07-20]
}
\bibitem{pagnia1999}{
    H Pagnia and F G\"artner.
    ``On the Impossibility of Fair Exchange without a Trusted Third Party.''
    Technical Report,
    Darmstadt University of Technology,
    1999-03-18.
    [online]
    \url{http://citeseerx.ist.psu.edu/viewdoc/download;jsessionid=522B5D1548F0A10C4C18E59C36986F55?doi=10.1.1.44.7863&rep=rep1&type=pdf}
    [retrieved 2021-02-12]
}
\bibitem{chaum1982}{
    D. Chaum.
    ``Blind Signatures for Untraceable Payments.''
    \textit{Advances in Cryptology: Proceedings of Crypto}
    \textbf{82}(3),
    pp. 199--203,
    1983.
    [online]
    \url{http://www.hit.bme.hu/~buttyan/courses/BMEVIHIM219/2009/Chaum.BlindSigForPayment.1982.PDF}
    [retrieved 2018-09-28]
}
\bibitem{chaum2021}{
    D Chaum, C Grothoff, and T Moser.
    ``How to issue a central bank digital currency.''
    Swiss National Bank Working Paper 3/2021,
    January 2021.
    [online]
    \url{https://www.snb.ch/n/mmr/reference/working_paper_2021_03/source/working_paper_2021_03.n.pdf}
    [retrieved 2021-02-25]
}
\bibitem{pboc2021}{
    Working Group on E-CNY Research and Development of the People's Bank of China.
    ``Progress of Research \& Development of E-CNY in China.''
    July 2021.
    [online]
    \url{http://www.pbc.gov.cn/en/3688110/3688172/4157443/4293696/2021071614584691871.pdf}
    [retreived 2021-07-16]
}
\bibitem{ap2018}{
    Associated Press.
    ``After 3 years, Greece ends limits on bank cash withdrawals.''
    2018-09-27.
    [online]
    \url{https://apnews.com/article/financial-markets-business-greece-europe-economy-57fcf1f4694a486aa4b7cd43095ae8ee}
    [retrieved 2022-12-20]
}
\end{thebibliography}
\end{document}